\documentclass[sigconf]{acmart}

\usepackage{booktabs}      %
\usepackage{tabularx}      %
\usepackage{array}         %
\usepackage{enumitem}      %
\usepackage{xcolor}        %
\definecolor{Bittersweet}{cmyk}{0,0.75,1,0.24}
\definecolor{RoyalBlue}{cmyk}{1,0.50,0,0}
\usepackage[normalem]{ulem}       %

\providecommand{\degree}{\mbox{\textdegree}}

\newif\ifrevision
\revisionfalse

\ifrevision
    \newcommand{\red}[1]{\textcolor{red}{#1}}
    \newcommand{\redsout}[1]{\textcolor{red}{\sout{#1}}}
\else
    \newcommand{\red}[1]{#1}
    \newcommand{\redsout}[1]{}
\fi

\newcommand{\system}{Map2Video}

\copyrightyear{2026}
\acmYear{2026}
\setcopyright{cc}
\setcctype{by}
\acmConference[UIST '26]{The 39th Annual ACM Symposium on User Interface Software and Technology}{November 02--05, 2026}{Detroit, MI, USA}
\acmBooktitle{The 39th Annual ACM Symposium on User Interface Software and Technology (UIST '26), November 02--05, 2026, Detroit, MI, USA}
\acmDOI{10.1145/3830398.3830538}
\acmISBN{979-8-4007-2856-3/2026/11}

\begin{document}

\title{\system{}: Guiding Real-World-Grounded AI Video Generation}

\author{Hye-Young Jo}
\orcid{0000-0003-3847-3420}
\affiliation{%
  \institution{University of Colorado Boulder}
  \city{Boulder, Colorado}
  \country{USA}}
\email{hye-young.jo@colorado.edu}
\authornote{This work was done while the first author was an intern at Fujitsu Research of America.}

\author{Mose Sakashita}
\orcid{0000-0003-4953-2027}
\affiliation{%
  \institution{Fujitsu Research of America}
  \city{Pittsburgh, Pennsylvania}
  \country{USA}}
\email{msakashita@fujitsu.com}

\author{Aditi Mishra}
\orcid{0009-0005-9628-8858}
\affiliation{%
  \institution{Fujitsu Research of America}
  \city{Pittsburgh, Pennsylvania}
  \country{USA}}
\email{amishra@fujitsu.com}

\author{Ryo Suzuki}
\orcid{0000-0003-3294-9555}
\affiliation{%
  \institution{University of Colorado Boulder}
  \city{Boulder, Colorado}
  \country{USA}}
\email{ryo.suzuki@colorado.edu}

\author{Koichiro Niinuma}
\orcid{0000-0001-8367-3988}
\affiliation{%
  \institution{Fujitsu Research of America}
  \city{Pittsburgh, Pennsylvania}
  \country{USA}}
\email{kniinuma@fujitsu.com}

\author{Aakar Gupta}
\orcid{0000-0001-6435-3583}
\affiliation{%
  \institution{Fujitsu Research of America}
  \city{Redmond, Washington}
  \country{USA}}
\email{agupta@fujitsu.com}

\renewcommand{\shortauthors}{Jo, et al.}

\begin{abstract}
AI video generation lowers barriers to video creation, but existing tools provide limited support for grounding content in real-world locations, restricting their use in applications such as film previsualization and scenario generation. In practice, users rely on verbose textual descriptions or reference images using on-site photos or street view imagery (SVI), which are cumbersome and unreliable across shots. Our formative study with five filmmakers revealed challenges in shot composition, character motion, and camera control when using SVI with an image-to-video tool. We present Map2Video, a system for guiding SVI-grounded AI video generation through map-based character trajectory annotation and direct manipulation of cameras and character masks defined in geodetic coordinates and projected onto panoramic screen coordinates. In an evaluation with twelve filmmakers, Map2Video provided stronger controllability for both scene replication and open-ended creative exploration, reduced cognitive effort, and improved spatial consistency across multiple shots compared to an image-to-video baseline.

\end{abstract}



\begin{teaserfigure}
\includegraphics[width=\textwidth]{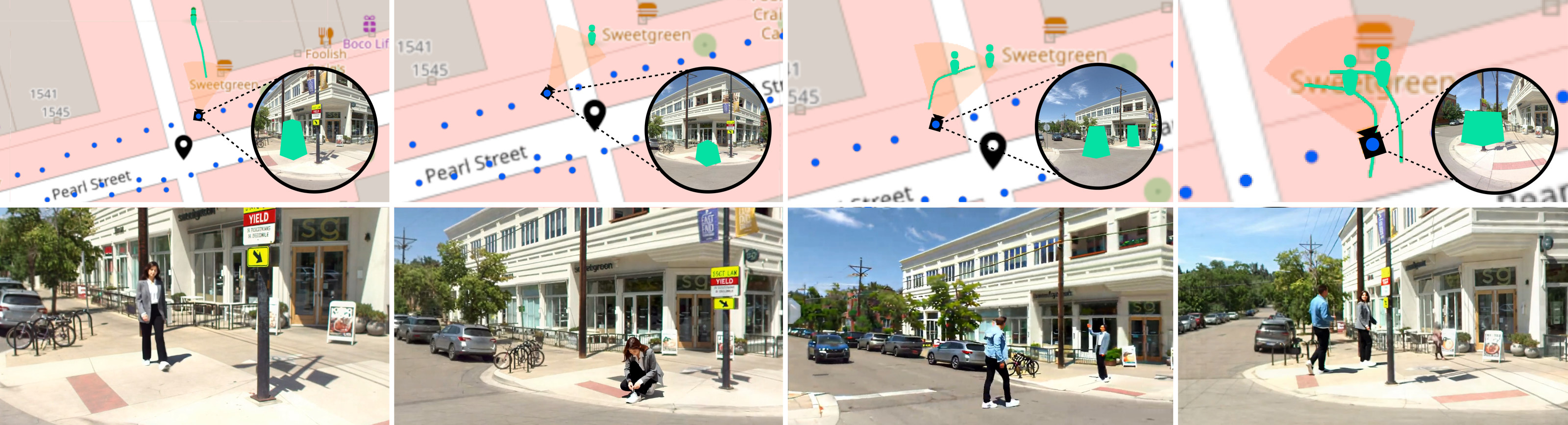}
\caption{\system{} helps users control video generation grounded in real-world locations by combining video inpainting with map-based annotation for character trajectory and masks and virtual camera framing on $360\degree$ street view imagery.}
\Description{Map2Video workflow showing map annotations, Street View framing with masks, and generated video frames with inserted characters.}
\label{fig:teaser}
\end{teaserfigure}

\maketitle

\section{Introduction}

Recent advances in text- and image-to-video generation have lowered barriers to video creation~\cite{brooks2024video, sun2024sora, yuan2024mora}, potentially reshaping filmmaking practices~\cite{zhang2025generative} and democratizing creativity. However, our survey with 34 AI filmmakers revealed recurring challenges in controlling AI video generation, such as difficulties with camera control and maintaining consistency across clips. Guided by these reported challenges and the portfolio examples shared by filmmakers, we examine how to support scenes in which creators need to control how a subject moves through a real-world environment and how the camera relates to that motion. Such scenarios arise in applications such as previsualizing scenes in specific locations before filming, prototyping scenes in public spaces that are difficult or expensive to access directly, and creating walking or driving shots in real-world locations.

To depict real-world locations in AI video generation, creators often rely on reference images~\footnote{\href{https://help.runwayml.com/hc/en-us/articles/40042718905875-Creating-with-Gen-4-Image-References}{Runway Gen-4 Image References Guidance}}, such as on-site photography or online search results (e.g., street view imagery)~\footnote{\href{https://curiousrefuge.com/blog/create-an-ai-film-in-any-location-with-google-maps}{Curious Refuge blog post on using street view imagery for AI video generation}}. While such references provide visual grounding, existing AI video generation interfaces offer limited support for specifying where the subject should appear, how it should move through the scene, and how the camera should relate to that motion. Meanwhile, maps and street view imagery provide spatial information tied to real-world locations, making them a promising foundation for specifying subject trajectories and planning camera behavior in grounded scenes.

To further investigate this opportunity, we used experience prototyping to examine a map- and street view imagery-driven workflow for grounded video creation in real-world locations. We created a lightweight proof-of-concept pipeline that combined a minimal map-based Street View Imagery (SVI)~\cite{biljecki2021street} retrieval interface with Runway, a popular AI video generation tool, and conducted a formative study with five filmmakers using this pipeline. Participants found the workflow promising for filmmaking, but they also identified four limitations: (1) difficulty specifying shot composition and subject placement via text prompts, (2) inconsistency in backgrounds caused by hallucination, (3) limited control over subject motion within the scene, and (4) lack of fine-grained control over camera movement.

Building on these insights, we developed \system{}, \red{a human-in-the-loop previsualization tool for video production in real-world locations that uses maps and street view imagery as an interactive authoring medium.} \system{} supports this workflow through map-based annotation of masks and trajectories, direct manipulation of virtual camera framing on 360\degree\ street view imagery, and video inpainting. Inspired by traditional filmmaking practices such as location scouting and on-location rehearsal, \system{} lets users refine subject placement, motion, and camera behavior in context before generation, supporting greater spatial control and continuity across shots.
\red{For example, an independent filmmaker planning a street scene can use the following six-step workflow:} 1) \textit{Location Scouting}, where users select scenes and backgrounds directly on a map; 2) \textit{Mask Positioning}, where proxy elements are placed into the street view scene and embedded in 3D space for later inpainting; 3) \textit{Movement Sketching}, where users draw trajectories to choreograph how actors or objects move across the scene, with synchronized updates in the camera view; 4) \textit{Camera Walkthrough}, where users refine \red{camera framing through panning, tilting, and zooming within selected SVI viewpoints}; 5) \textit{Scene Prompting}, where users provide natural language descriptions of characters’ appearances and actions; and 6) \textit{Video Inpainting}, where the system combines these inputs to generate \red{a reference clip before committing time and cost to on-site filming}. We implemented \system{} using Unity and ComfyUI with the VACE video generation model, along with OpenStreetMap and Mapillary for map and street view imagery.

We evaluated \system{} with 12 filmmakers across replication and open-ended tasks. In the replication task, \system{} enabled \red{participants to complete the task $1.2\times$ faster by reducing prompt-based compositional trial and error} while achieving higher perceived controllability than an image-to-video baseline. Across tasks, participants reported lower cognitive effort, higher usability, greater creative controllability, and higher perceived spatial consistency. These results suggest that \system{} better supports controllable, location-grounded video authoring for subject motion and camera planning in real-world scenes.

In summary, our main contributions are:
\begin{enumerate}[nosep,leftmargin=2.0em]
    \item An experience prototyping study ($N=5$), informed by a survey of AI filmmaking practices ($N=34$), that identified key challenges and interaction needs for grounded video creation in real-world scenes, particularly around subject movement and camera control.
    \item \red{\system{}, a human-in-the-loop AI video authoring pipeline that uses maps and street view imagery as an interactive authoring medium, supporting visible and editable control of subject placement, motion trajectories, and camera framing through map-based annotation, direct manipulation on SVI, and video inpainting.}
    \item \red{A user evaluation ($N=12$) comparing \system{} to an image-to-video baseline, demonstrating higher controllability, lower cognitive effort, higher perceived spatial consistency across both tasks, and $1.2\times$ faster task completion in the replication task.}
\end{enumerate}

\section{Related Work}

\subsection{Generative Video and Grounding}

\red{Recent text-to-video models~\cite{brooks2024video, veo, runway, sun2024sora} have lowered barriers to video production and influenced emerging filmmaking practices~\cite{zhang2025generative, halperin2025underground, anderson2025making}. Prior work improves explicit control over motion and camera behavior~\cite{yang2024direct, wang2024motionctrl}, temporal stability and identity preservation~\cite{ren2024consisti2v, bai2025syncammaster, chen2024videocrafter2}, and fine-grained conditioning through pose sequences and segmentation masks~\cite{hu2024animate, jiang2025vace}. Commercial tools further provide camera rig and lens-level control within 3D environments~\cite{higgsfield2026cinema}. These works improve controllability and output consistency during video generation.}

\red{Recent work also grounds video generation in real-world spatial data. \textit{Streetscapes}~\cite{deng2024streetscapes} uses map layouts, \textit{StreetCrafter}~\cite{yan2025streetcrafter} uses LiDAR-based geometric signals for camera control, \textit{DreamDrive}~\cite{mao2025dreamdrive} generates 4D driving scenes from street view imagery, and \textit{Seoul World Model}~\cite{seo2026grounding} uses retrieved street view imagery to ground urban simulation. These approaches primarily use spatial data as model-level conditioning signals or scene representations.}

\red{\system{} instead exposes maps and street view imagery as user-facing authoring materials, rather than using them only as model-level inputs.}

\subsection{Video Authoring Interfaces}

\red{Prior HCI work supports video authoring through three broad approaches. Some tools edit, explore, and adapt existing footage using transcripts, annotations, signals, and language-based interaction~\cite{huber2019b, huh2023avscript, moorthy2020gazed, chang2021rubyslippers, leake2017computational, truong2016quickcut, leake2024chunkyedit, tilekbay2024expressedit, wang2024lave, sandoval2025editduet, li2025shots, huh2025videodiff, lin2022videomap, ruangrotsakun2023viva, kim2023surch, pavel2015sceneskim, murakami2024swapvid, yang2025vidstr}. Others repurpose existing media into highlights, summaries, news, and teaser videos~\cite{wang2022record, barua2025lotus, van2024making, lin2024videogenic, kim2025generating, wang2024reelframer, wang2024podreels}. A third group creates videos from text, transcripts, scripts, documents, web pages, and other source content~\cite{wang2019write, xia2020crosscast, xia2020crosspower, leake2020generating, chi2022synthesis, chi2020automatic, chi2021automatic}. Across these approaches, authoring begins with existing footage or other source content.}

\red{Moving beyond source-based workflows, recent works make generated video itself the object of authoring. \textit{VideOrigami}~\cite{cao2025compositional} and \textit{Doki}~\cite{liu2026text} organize generation through timelines, canvases, narrative structures, and text-based views. This direction also spans music video production, animation, and map animation~\cite{liu2023generative, lee2025mvprompt, jahanlou2022katika, liu2025logomotion, gunturu2025mapstory}. For cinematic authoring, \textit{DreamCinema}~\cite{chen2024dreamcinema} leverages existing film scenes to transfer camera trajectories and character motion to generated video, while \textit{CineVerse}~\cite{phung2025cineverse} generates coherent multi-shot keyframes from high-level scene descriptions.}

\red{\system{} extends this landscape by adding location-grounded spatial authoring  using maps and street view imagery to shape generated scenes in specific real-world locations.}

\subsection{Previsualization and Map-Based Storytelling}

\red{HCI research has explored previsualization tools for cinematic composition and planning. \textit{CollageVis}~\cite{jo2024collagevis} and \textit{CinemAssist}~\cite{he2024interactive} support rapid previews and composition guidance, while other systems generate director-style cut scenes or provide previs on virtual 3D stages~\cite{evin2022cine, chen2024cinepregen}. Script-to-previs pipelines further support controllable cameras and virtual actors~\cite{zhu2023moviefactory, wang2025script2screen, kim2021asap}. These approaches primarily rely on pre-authored assets and virtual environments.}

\red{Location-aware and map-based systems connect narratives and visual media to physical places~\cite{belz2024story, kim2023dynamic, li2023location}. Other systems support camera-path planning and viewpoint transitions in geographic visualizations~\cite{shin2023space, li2023geocamera}. This area also includes map animations~\cite{gunturu2025mapstory}, embodied data narratives~\cite{jain2025strollytelling}, navigation of 360\degree{} media~\cite{lee2024viewer2explorer}, and geospatial creative workflows using platforms like \textit{Cesium}~\cite{cesium2023}.}

\red{\system{} bridges previsualization and map-based authoring by using maps and street view imagery as direct materials for location-grounded scene creation.}

\section{Formative Study}

\subsection{Survey of AI Filmmaking Practices}

\red{Building on prior work documenting emerging AI filmmaking practices~\cite{zhang2025generative}, we conducted an online survey with 34 filmmakers to examine current tool use, workflows, and challenges in AI-assisted filmmaking. Participants were recruited through online AI filmmaking communities. The full method and broader findings are reported in Appendix Section~\ref{subsection:survey}. Here, we focus on the findings that directly informed our design direction, including tool use, location-based reference practices, and challenges in spatial control.}

\red{Most participants (29/34) used multiple AI tools rather than relying on a single platform. Runway was mentioned most frequently (29/34), followed by Kling (18/34), Veo (12/34), and ComfyUI (11/34). Their responses indicated diverse strategies for preparing reference images for video generation, including real-world photographs, generated images, and virtual environments. Participants also described using AI filmmaking to fill gaps such as location footage, animation, or visual effects, and to visualize ideas that would be difficult or costly to film conventionally.}

\red{Participants reported difficulties with visual and spatial consistency (17), prompt adherence (12), character placement (11), composition and camera movement (9), and character actions (3). Achieving coherent, high-quality outputs often required extensive iteration, manual correction, and post-production. These findings motivated us to explore more direct spatial controls for subject placement, motion, and camera behavior in location-grounded scenes.}

\subsection{SVI-Driven AI Video Generation}

\red{The survey showed that filmmakers already used external visual references and encountered recurring difficulties in controlling generated videos. This led us to consider whether real-world location imagery could play a more active role in the authoring process rather than serving only as a reference input. We therefore explored SVI-driven AI video generation, in which maps, street-level panoramas from services such as Google Streetview and Mapillary, and map annotations provide a spatial workspace and spatial references for video diffusion models (Figure~\ref{fig:teaser}).}

\red{This direction appeared promising because it connects video generation with familiar filmmaking practices such as location scouting and previsualization while directly producing video outputs. Reusing imagery from the same location may also help preserve scene geometry across shots. SVI is particularly suitable for outdoor, location-grounded authoring because of its coverage of public spaces. However, its limited indoor coverage and discrete viewpoints constrain indoor authoring, smooth lateral motion, and parallax. These limitations could be addressed by complementing SVI with other forms of spatial capture, 3D reconstruction, and novel-view synthesis. Indoor locations could be captured and reconstructed using techniques such as NeRF and 4D Gaussian splatting, while \textit{Streetscapes}~\cite{deng2024streetscapes} and \textit{StreetCrafter}~\cite{yan2025streetcrafter} suggest possible paths toward smoother transitions and more flexible camera trajectories in outdoor scenes.}

\subsection{Experience Prototyping}\label{section:formative}

\red{To investigate how filmmakers might use such a workflow and what forms of control it would require, we conducted experience prototyping~\cite{buchenau2000experience, camburn2017design} with a lightweight proof-of-concept pipeline. By simulating the end-to-end experience of location scouting, selecting panoramas, and generating AI videos, we examined how participants adapted to the workflow, where they encountered difficulties, and what interactions they desired.}

\begin{figure}[h]
    \centering
    \includegraphics[width=\linewidth]{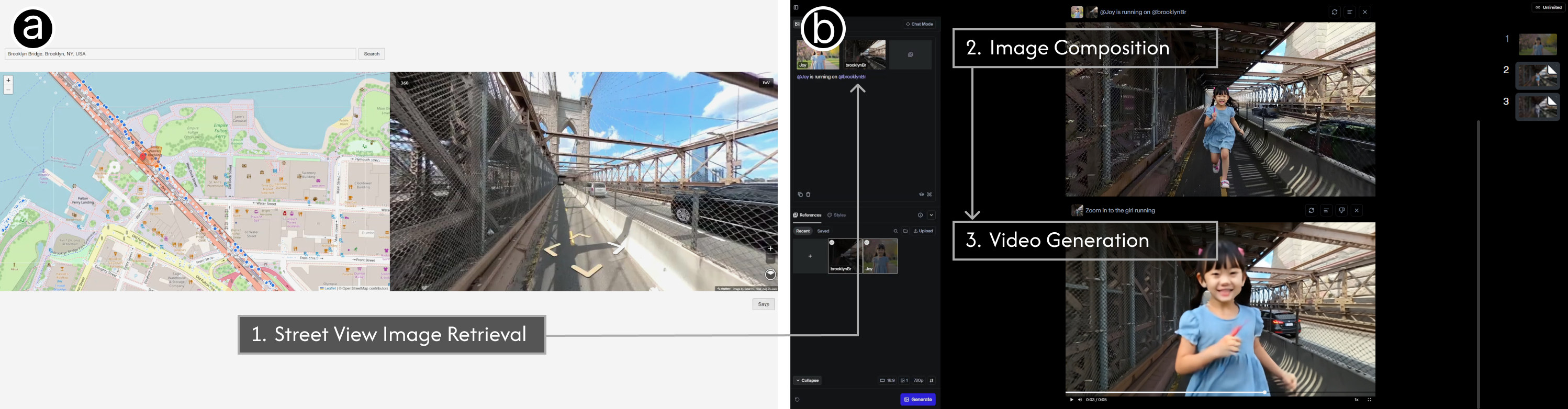}
    \caption{Proof-of-concept pipeline of SVI-driven AI video generation: \textcircled{a} SVI search, \textcircled{b} AI video generation (Runway)}
    \label{fig:formative-study}
    \Description{Proof-of-concept pipeline showing map and streetview input on the left, and image composition and video generation on the right.}
\end{figure}

\subsubsection{\textbf{Proof-of-concept Pipeline}}
    \begin{enumerate}[nosep,leftmargin=2.0em]
    \item \textbf{Map and Street View Imagery Exploration:} a map-based street view imagery search tool~\footnote{\href{https://map2streetview.vercel.app/}{https://map2streetview.vercel.app/}} (OpenStreetMap + Mapillary) allows users to browse $360\degree$ panoramas, set yaw and pitch, and save viewpoints as reference backgrounds.
    \item \textbf{Video Generation with Reference Images:} participants upload the saved backgrounds to a popular AI video tool (Runway~\footnote{\href{https://app.runwayml.com/}{https://app.runwayml.com/}}), add a character reference image, and write prompts for actions and camera motions such as pan, tilt, and dolly.
\end{enumerate}
We intentionally kept the prototype minimal and manually \red{transferred} between tools, similar to how practitioners currently work.

\subsubsection{\textbf{Participants}}
We recruited 5 of 34 survey participants (1 female, 4 male) who had mentioned challenges related to \textit{``spatial,''} \textit{``environments,''} or \textit{``geography''} in their responses, to gather deeper insights into their experiences and strategies for managing spatial context in AI-assisted filmmaking. All participants had prior experience in both traditional filmmaking ($M = 10.8$, $SD = 11.5$ years) and AI-assisted filmmaking ($M = 10.2$, $SD = 2.5$ months).

\subsubsection{\textbf{Protocol}}
The study included two parts: 1) a semi-structured interview about challenges in spatially grounded AI video creation, and 2) an exploratory experience prototyping session using the proof-of-concept pipeline. During prototyping, participants shared their screens and thought aloud while creating 3--5 videos within a single location. They searched for candidate locations, selected $360\degree$ panoramas, saved viewpoints as reference backgrounds, and used them in Runway to generate videos (Figure~\ref{fig:formative-study}).

\begin{figure}[hbt]
    \centering
    \includegraphics[width=\linewidth]{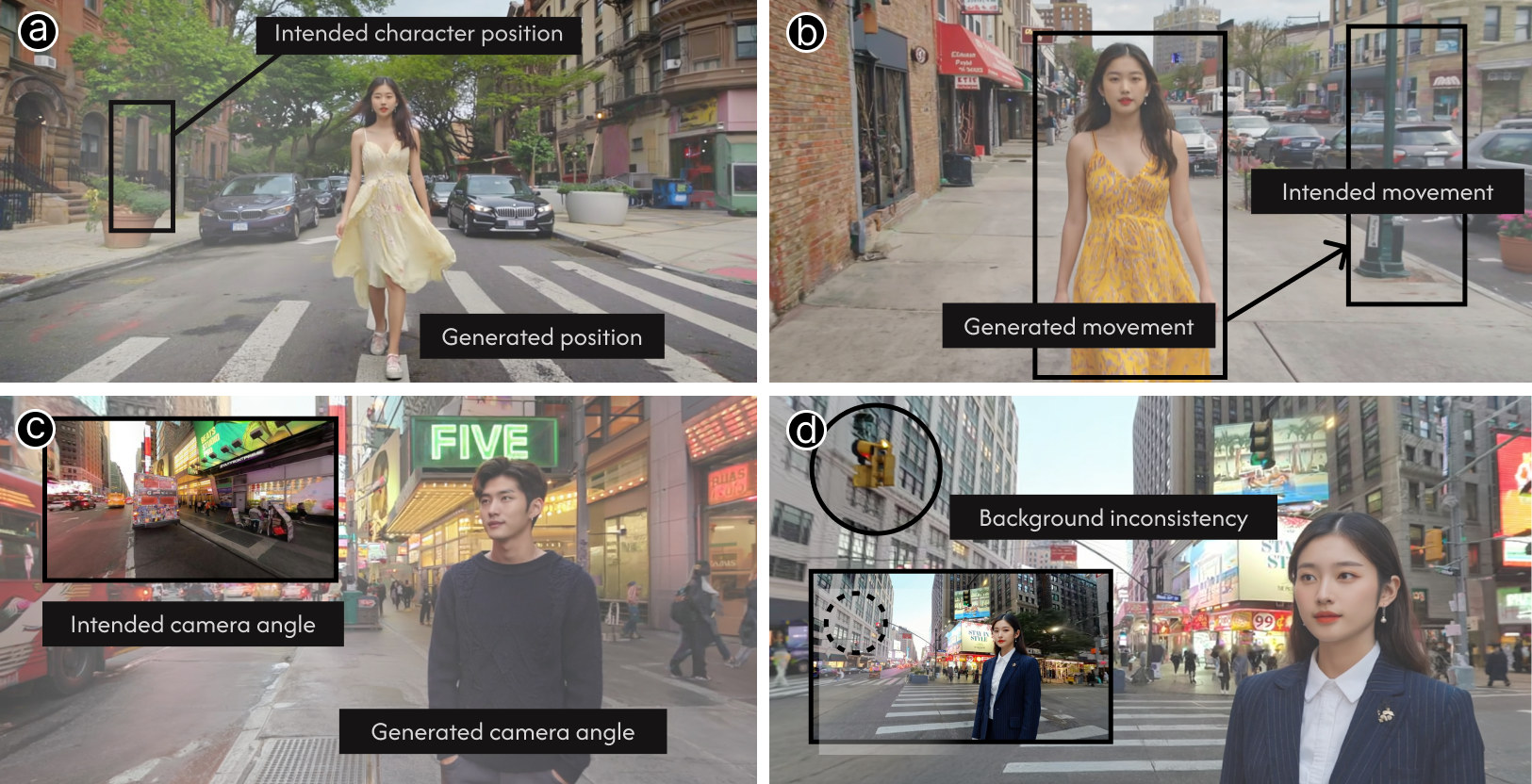}
    \caption{Challenges in spatially grounded video generation.}
    \label{fig:spatial-continuity-challenges}
    \Description{Examples of spatial continuity challenges, including mismatched character position, movement, camera angle, and background consistency.}
\end{figure}

\subsubsection{\textbf{Challenges and Desired Interactions}}
Overall, participants responded positively to the SVI-driven workflow, noting that virtual exploration of candidate locations helped them reason more concretely about scene layout and camera setup (see Appendix Figure~\ref{fig:formative-user-artifact-example} for an example participant-created sequence). Despite these benefits, participants described recurring challenges in controlling generated videos within this workflow (Figure~\ref{fig:spatial-continuity-challenges}).

\subsubsection*{\textbf{Placing Characters in Space}}
Specifying where a character should appear via text prompts proved unreliable across models (Figure~\ref{fig:spatial-continuity-challenges}.\textcircled{a}). Participants tried screen-direction terms (e.g., ``screen left,'' ``foreground''; P5) or alignment with reference footage (P3), yet outcomes varied, forcing trial and error and compromising framing (P1, P4, P5). They preferred to place characters directly on a map or within the frame, visually anchoring position and depth so that intent was unambiguous and more consistent across models (P5).

\subsubsection*{\textbf{Directing Character Motion}}
Controlling character motion was also limited (Figure~\ref{fig:spatial-continuity-challenges}.\textcircled{b}). Participants aligned the last frame of one clip with the first of the next (P1) or trimmed away erratic movement and physical violations (P3, P5). They envisioned path-based control, such as sketching trajectories and key poses on the map or in the frame, so the system could generate transitions that preserved intended spatial relationships across shots (P1, P2, P5).

\subsubsection*{\textbf{Exploring Camera Movement}}
Camera motion often collapsed to generic angles despite reference images and detailed prompts (P1, P5) (Figure~\ref{fig:spatial-continuity-challenges}.\textcircled{c}). Participants generated many takes and curated usable clips. Some maintained prompt libraries (P3), while others preferred tools that parsed cinematographic terms more effectively (P5). They wanted free navigation within $360\degree$ panoramas and the ability to set multiple camera positions on a map to preview pans, tilts, and dollies before committing to generation (P2, P3, P5).

\subsubsection*{\textbf{Preserving Background Continuity}}
While the SVI-driven workflow helped maintain background continuity, participants still observed inconsistencies across multi-shot sequences, such as hallucinated or shifting props (P5), uneven environmental effects (P5), and disappearing or altered details (P2, P3) (Figure~\ref{fig:spatial-continuity-challenges}.\textcircled{d}). Workarounds included choosing simple backgrounds, using shallow depth of field, and hiding cuts with close-ups (P2, P4, P5). They envisioned selective editing or automatic continuity auditing that could flag inter-shot background differences for correction (P2, P5).

\subsubsection*{\textbf{Key Takeaway}}
Across these themes, participants converged on the need for more directly controllable interfaces for grounded video authoring, particularly for subject placement, motion path sketching, and camera preview in map and panorama views.

\section{\system{} System}

Based on the formative study, we developed \system{}, a map-based authoring tool for SVI-grounded video generation. \system{} uses street view imagery to support direct control over subject placement, subject motion, and camera behavior before rendering.

\begin{figure}[H]
    \centering
    \includegraphics[width=\linewidth]{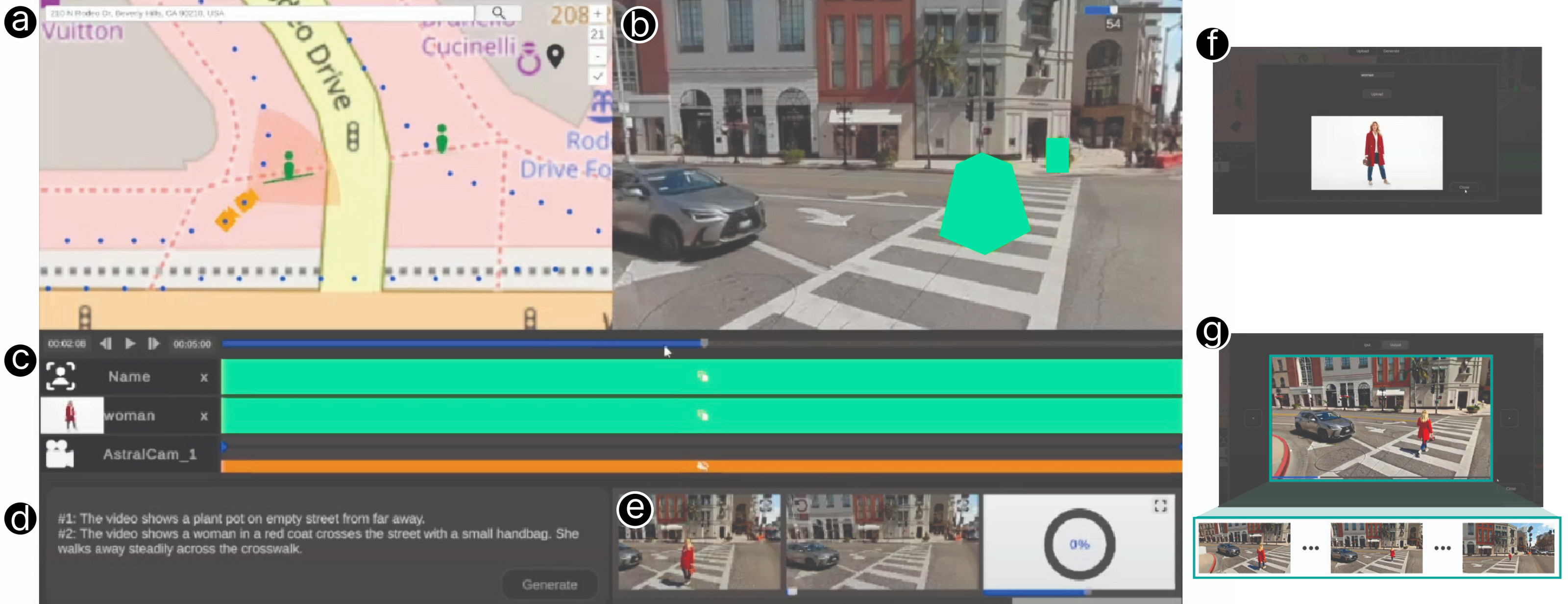}
    \caption{\system{} interface:
    \textcircled{a} Map Panel,
    \textcircled{b} SVI Panel,
    \textcircled{c} Timeline Panel,
    \textcircled{d} Prompt Panel,
    \textcircled{e} Render Queue Panel,
    \textcircled{f} Reference Image Window, and
    \textcircled{g} Video Player Window.}
    \label{fig:interface}
    \Description{Map2Video interface showing panels for map, Street View, timeline, prompt input, render queue, and windows for reference image, and video playback.}
\end{figure}

\begin{figure*}[h]
    \centering
    \includegraphics[width=\textwidth]{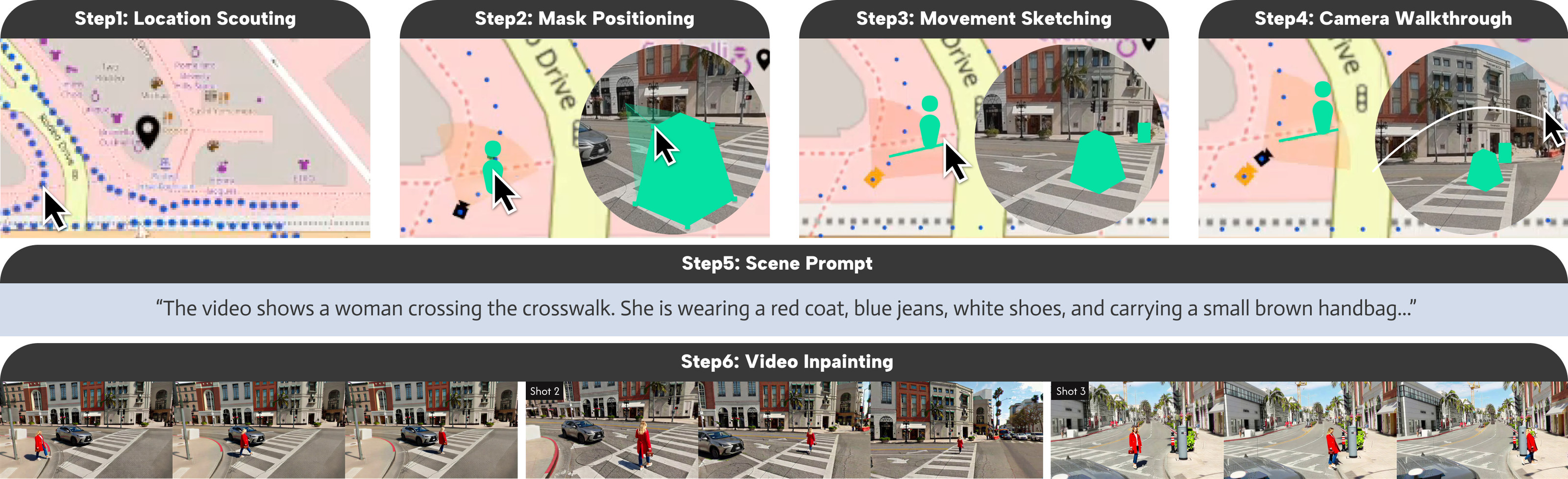}
    \caption{\system{} system walkthrough.}
    \label{fig:walkthrough}
    \Description{Map2Video system walkthrough showing steps from location scouting and mask placement to movement sketching, camera walkthrough, prompt entry, and final inpainted video output.}
\end{figure*}

\subsection{System Walkthrough}
As shown in Figure~\ref{fig:walkthrough}, \system{} combines map-based planning and video inpainting through six steps.

\subsubsection*{\textbf{Step 1. Location Scouting:}}
Users first choose a location in the Map Panel (Figure~\ref{fig:interface}\textcircled{a}) and select a street view image as the scene background. The SVI Panel (Figure~\ref{fig:interface}\textcircled{b}) previews the selected view and allows users to adjust framing, such as view angle and field of view, while keeping the shot grounded in the chosen location. \red{Although the available SVI constrains the initial background composition, \system{} treats the selected view as editable input rather than a fixed background. The workflow can refine it before subject inpainting, for example by removing unwanted pedestrians and upscaling low-resolution imagery (Figure~\ref{fig:upscaling-multimask}).}

\subsubsection*{\textbf{Step 2. Mask Positioning:}}
Users place a green \red{polygonal} mask, which serves as \red{a} subject proxy, in the Map Panel (Figure~\ref{fig:interface}\textcircled{a}). The system projects this proxy into the SVI Panel (Figure~\ref{fig:interface}\textcircled{b}) as a 2D card anchored to the scene. This lets users determine where a subject should appear in the grounded environment before generation.

\subsubsection*{\textbf{Step 3. Movement Sketching:}}
Users sketch the movement path of the mask in the Map Panel (Figure~\ref{fig:interface}\textcircled{a}), and the corresponding motion is synchronized in the SVI Panel (Figure~\ref{fig:interface}\textcircled{b}). This allows them to specify how the subject moves through the scene while previewing the motion in context.

\subsubsection*{\textbf{Step 4. Camera Walkthrough:}}
Users define \red{camera framing over time in the SVI Panel (Figure~\ref{fig:interface}\textcircled{b}) by adjusting pan, tilt, and zoom within selected SVI viewpoints.} The Timeline Panel (Figure~\ref{fig:interface}\textcircled{c}) synchronizes subject motion and camera actions, allowing users to preview and refine the shot structure in relation to the scene. \red{As a proof of concept, the current prototype uses the discrete viewpoints available in SVI and does not integrate continuous 3D camera translation between them. Future implementations could extend this capability by reconstructing or synthesizing intermediate views along a camera trajectory using techniques demonstrated by \textit{Streetscapes}~\cite{deng2024streetscapes} and \textit{StreetCrafter}~\cite{yan2025streetcrafter}.}

\subsubsection*{\textbf{Step 5. Scene Prompt:}}
Once users are satisfied with the spatial setup, motion, and camera plan, they describe the target subject or event in the Prompt Panel (Figure~\ref{fig:interface}\textcircled{d}) and submit the shot to the Render Queue (Figure~\ref{fig:interface}\textcircled{e}). Users can optionally provide a reference image in the Reference Image Window (Figure~\ref{fig:interface}\textcircled{f}) to guide subject appearance across iterations or shots. Otherwise, the system generates the subject from the prompt.

\begin{figure}[h]
    \centering
    \includegraphics[width=\linewidth]{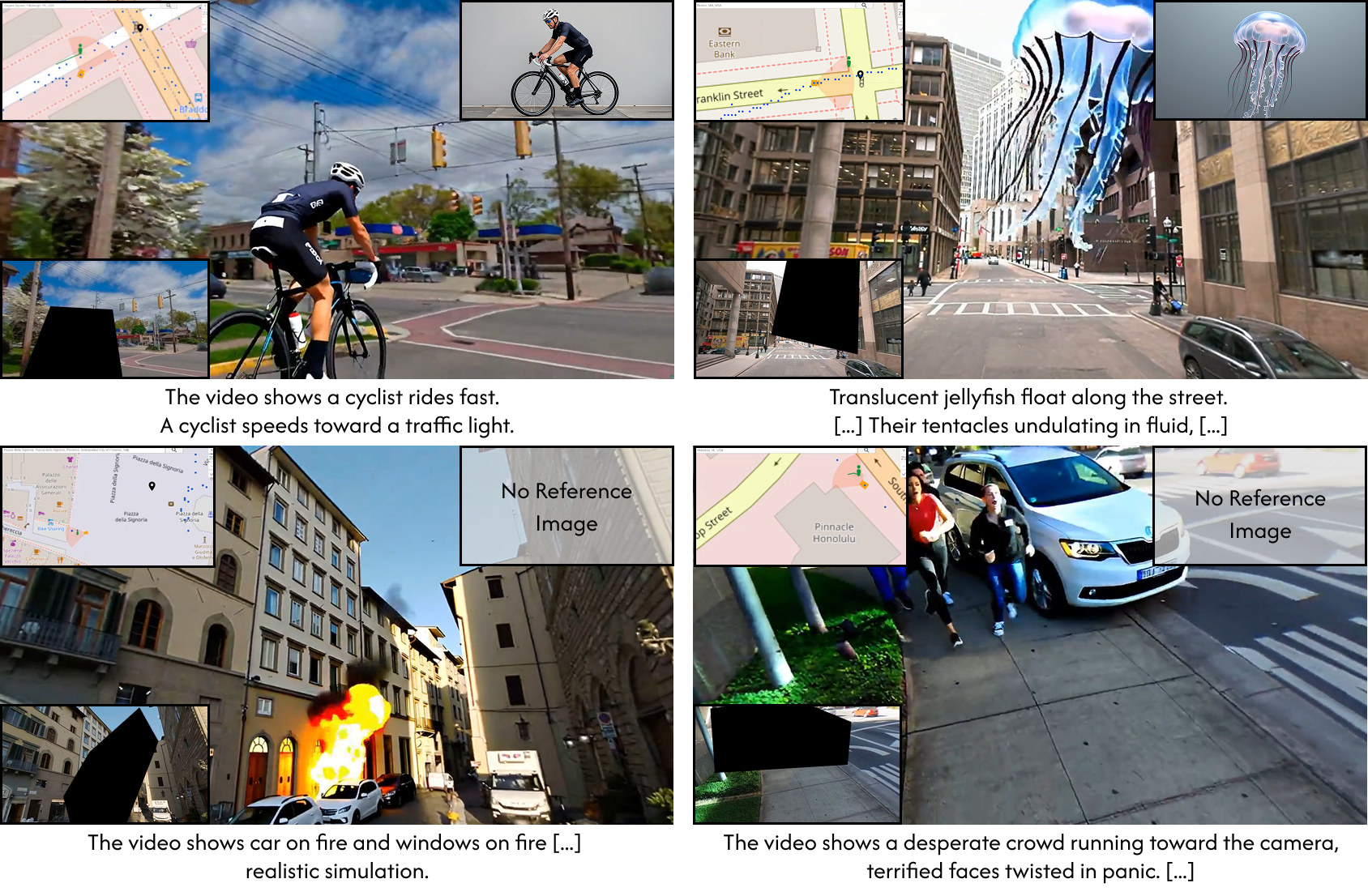}
    \caption{Various inpainted subjects created with or without reference images, such as a cyclist, jellyfish, fire, and a crowd.}
    \label{fig:various-subjects}
    \Description{Inpainted subjects generated with or without reference images, including a cyclist, floating jellyfish, fire effects, and a running crowd.}
\end{figure}

\begin{figure*}[h]
    \centering
    \includegraphics[width=\textwidth]{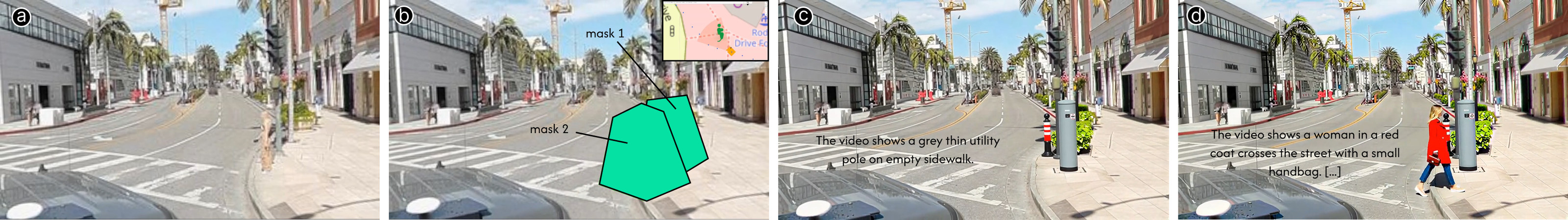}
    \caption{\red{Background refinement and multi-mask inpainting process.} \textcircled{a} original image, \textcircled{b} masked image, \textcircled{c} pedestrian removal and background upscaling, and \textcircled{d} character composition.}
    \label{fig:upscaling-multimask}
    \Description{Multi-mask inpainting process showing original image, masked image, pedestrian removal with background upscaling, and new character composition.}
\end{figure*}

\subsubsection*{\textbf{Step 6. Video Inpainting:}}
Finally, the system generates the video by inpainting only the masked region based on the specified scene, motion, camera, and prompt. This supports iterative refinement of the subject while preserving the surrounding scene. Outputs can be reviewed in the Video Player (Figure~\ref{fig:interface}\textcircled{g}).

\red{Users can adjust mask scale to create grounded scene elements, including realistic characters, environmental events such as fire, surreal objects such as flying jellyfish, and crowds (Figure~\ref{fig:various-subjects}). The system also supports multiple masks, processed sequentially at the cost of increased generation time (Figure~\ref{fig:upscaling-multimask}).}

\subsection{Implementation}

\begin{figure*}[h]
    \centering
    \includegraphics[width=\textwidth]{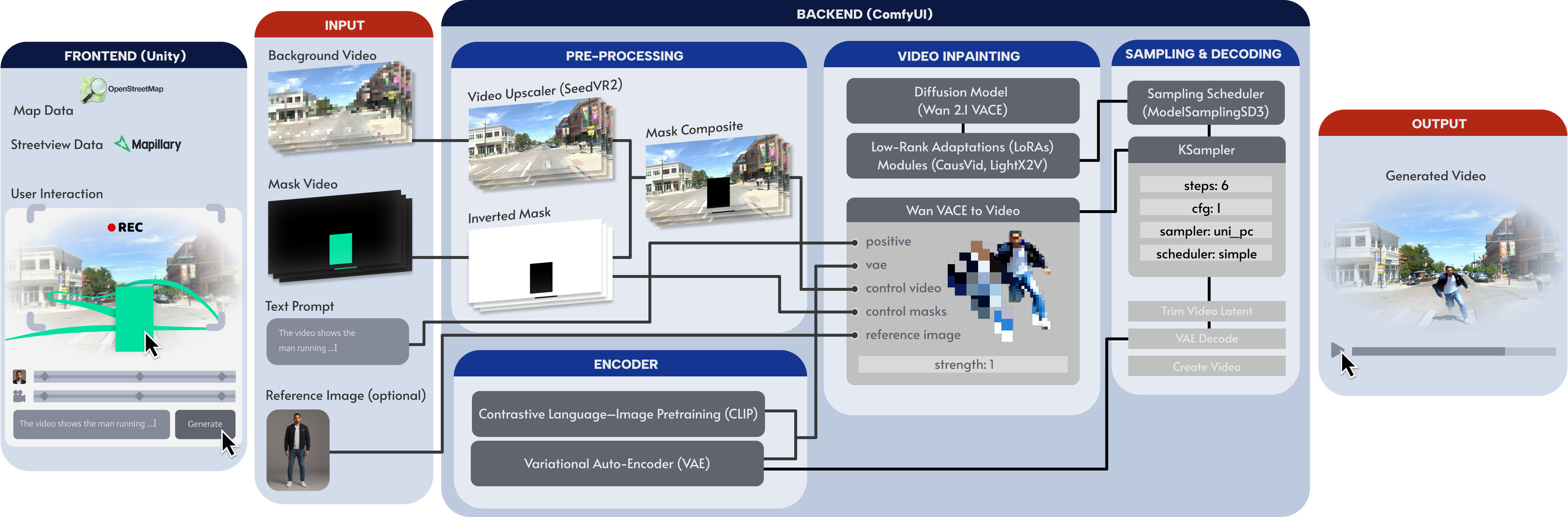}
    \caption{\system{} system architecture.}
    \label{fig:system-architecture}
    \Description{Map2Video system architecture showing inputs (prompt, image, mask), preprocessing, diffusion-based inpainting, and final video output.}
\end{figure*}

As shown in Figure~\ref{fig:system-architecture}, \system{} is implemented with Unity as the frontend and ComfyUI, an open-source node-based interface for diffusion models, as the backend. The ComfyUI workflow runs on a remote server equipped with an NVIDIA H100 NVL GPU.

Based on user interaction in Unity, ComfyUI takes four inputs to generate the video: (1) a background video, consisting of a street view imagery sequence with camera movement; (2) a mask video, defining a geographically fixed mask position and movement aligned with the camera work; (3) text prompts describing the intended scene; and (4) an optional reference image, which users can upload or generate to guide the masked content. ComfyUI then upscales the background video and inpaints only the masked region. Each sequence is limited to five seconds by the Wan2.1 model~\cite{wan2025}, and longer videos can be created by stitching multiple clips together~\cite{li2026stable}.

\subsubsection{\textbf{Map Interaction and Mask Placement}}
We use OpenStreetMap for the map and Mapillary for street view imagery. \red{Unless otherwise noted, all street-level imagery in this paper is courtesy of Mapillary, and all map data is copyrighted by OpenStreetMap contributors.} To place a subject in the panoramic view, we project the subject's map coordinates into the camera view. We approximate the local area with a tangent plane (East--North--Up) centered at the camera, compute the subject's relative azimuth and elevation, and map these angles to pixel coordinates using a pinhole camera model. \red{The mask is initialized as a rectangular polygon with default real-world dimensions of $1.0 \times 2.0\,\mathrm{m}$. Its position and size in the SVI Panel are computed from the map and SVI geometry, so the projected mask appears smaller as its distance from the camera increases. Users can resize or reshape the mask by dragging its vertices and can add new vertices by hovering over an edge and inserting a control point.} Details are provided in Appendix Table~\ref{tab:actorplacement} for reproducibility.

\subsubsection{\textbf{Video Inpainting Workflow}}
Our video inpainting workflow builds on the default VACE pipeline in the ComfyUI documentation~\footnote{\url{https://docs.comfy.org/tutorials/video/wan/vace/}} and adds nodes for street-view-based video authoring. VACE is a unified video foundation model for text-to-video, image-to-video, and masked video-to-video tasks. It builds on the Wan2.1 and integrates reference and mask inputs through a video condition unit and context adapter, enabling flexible video generation~\cite{jiang2025vace}.

\subsubsection*{\textbf{Pre-processing}}
To address the low resolution of $360\degree$ Mapillary imagery, we use SeedVR~\cite{wang2025seedvr2} to upscale low-resolution $360\degree$ Mapillary imagery. The enhanced background is then combined with an inverted mask and fed into the inpainting module.

\subsubsection*{\textbf{Encoder}}
We use \texttt{umt5-xxl fp8 e4m3fn scaled (wan type)} to encode text prompts into conditioning vectors for diffusion. We then use the \texttt{Wan 2.1 VAE} to encode frames into latent space and decode them back into RGB frames, reducing computation.

\subsubsection*{\textbf{Video Inpainting}}
Inpainting is handled by the \texttt{Wan VACE To Video} node. The upscaled background with an inverted alpha mask constrains edits to masked regions while preserving the grounded background and camera motion. We use \texttt{Wan2.1 VACE 14B fp16 (weights dtype: fp8 e4m3fn fast)} for diffusion model.

We also use two LoRA modules: CausVid~\cite{yin2025causvid} to accelerate convergence and LightX2V~\cite{lightx2v} to improve temporal stability. In our environment, the workflow takes about 50 seconds to load the model, around 3 minutes to sample a 5-second 1280$\times$720 video at 16 fps, and another 40 seconds to render and download the video from the server.

Although optional, a reference image helps maintain subject appearance. Users can upload or generate one in the reference image window (Figure~\ref{fig:interface}.\textcircled{f}). If a user chooses to generate a character instead of uploading a subject image, we run a lightweight workflow using the Unet Loader with a GGUF-formatted model (\texttt{Wan2.1 14B VACE Q4 K M.gguf}). We also apply the \texttt{Wan14B Realism Boost T2V LoRA} so the generated character better matches real-world street view imagery.

\subsubsection*{\textbf{Sampling and Decoding Strategy}}
For inference, we use the \texttt{Model Sampling SD3}~\cite{patrick2024scaling} scheduler followed by \texttt{KSampler} in ComfyUI. The \texttt{SD3} shift parameter (set to 5) balances coarse structure and detail. \texttt{KSampler} runs six steps (guidance scale 1, \texttt{uni\_pc}, \texttt{simple} scheduler, strength 1), prioritizing fast generation while maintaining background fidelity.

\section{User Evaluation}
We evaluated \system{} with 12 filmmakers to examine whether it provides greater control than an image-to-video baseline over subject placement, subject motion, and camera behavior in SVI-grounded video generation.

\begin{figure}[h]
    \centering
    \includegraphics[width=\columnwidth]{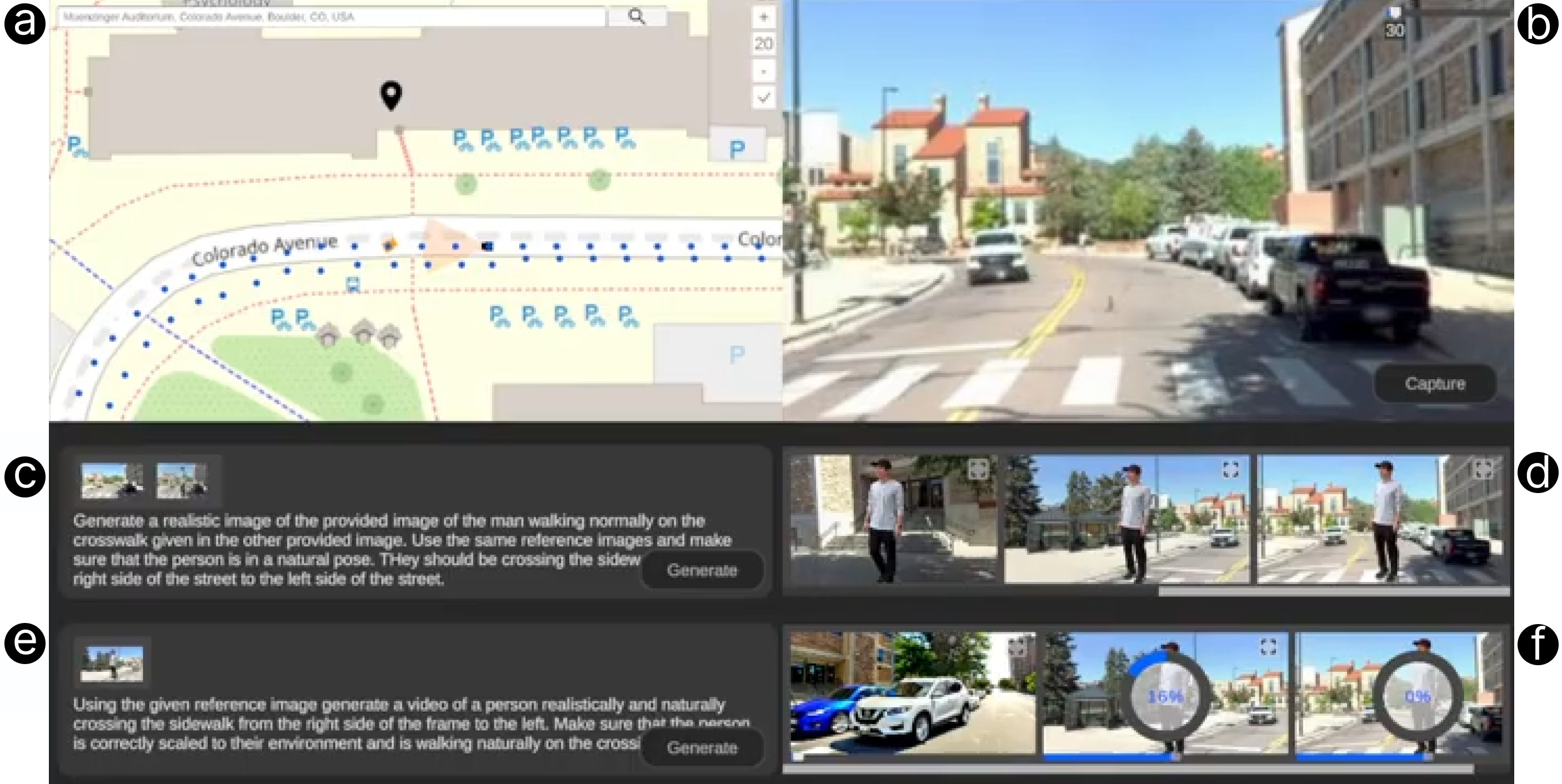}
    \caption{Baseline interface:
    \textcircled{a} Map Panel,
    \textcircled{b} Street View Imagery Panel,
    \textcircled{c} Image Prompt Panel,
    \textcircled{d} Video Prompt Panel,
    \textcircled{e} Image Render Queue Panel, and
    \textcircled{f} Video Render Queue Panel.}
    \label{fig:baseline}
    \Description{Baseline interface showing panels for map, Street View, image prompt input, image render queue, video prompt input, video render queue, and windows for reference image, and video playback.}
\end{figure}

\subsection{Method}
We used a within-subjects design in which participants completed two video authoring tasks using both \system{} and an image-to-video baseline.

\subsubsection{\textbf{Participants}}

\red{We use \textit{`filmmaker`} broadly to refer to people who create film or cinematic video projects and make directorial decisions about story, blocking, composition, camera, and editing. This definition includes formally trained, self-taught, and cross-disciplinary pathways~\cite{rabiger2020directing}. Recruitment required at least one year of such filmmaking experience.}

We recruited 12 participants (ages 20--46; $M=28.00$, $SD=7.47$; 5 female, 7 male). Their filmmaking experience ranged from 1--12 years ($M=5.38$, $SD=3.18$). \red{Participants included five film students, two film-school graduates with professional film-industry experience, three directors with fine art backgrounds, and two cinematic content creators who had produced short films and cinematic documentaries.} All had prior experience with AI video tools. Five reported using them actively in their own work, while the others had used them more casually for experimentation.

\begin{figure}[h]
    \centering
    \includegraphics[width=\columnwidth]{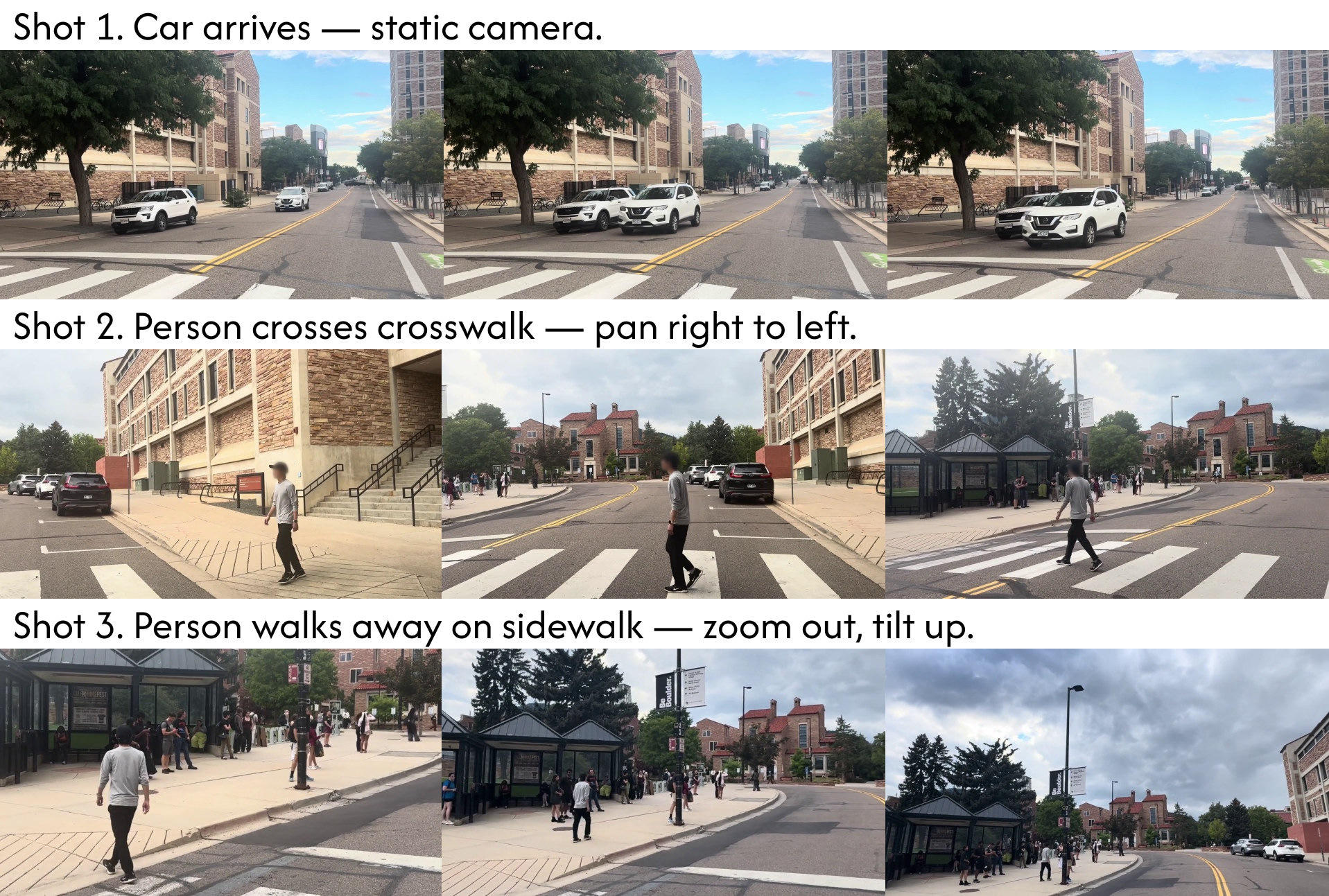}
    \caption{Replication task material.}
    \label{fig:replication-task}
    \Description{Replication task with three shots: a car arriving in a static frame, a person crossing the crosswalk with camera pan, and the person walking away with zoom out and tilt up.}
\end{figure}

\subsubsection{\textbf{Conditions}}

\red{We compared \system{} with a commonly accessible Runway-style image-to-video workflow, informed by Runway being the most frequently reported AI video tool in our survey. The baseline represented a conventional image-to-video workflow.}

\begin{enumerate}[leftmargin=*]
    \item \textbf{Baseline} (Figure~\ref{fig:baseline}): An image-to-video (I2V) tool with map-based street view retrieval. Participants selected a background and subject image to create a composite reference image for video generation. We added automatic retrieval to avoid manual download and upload.

    \item \textbf{\system{}} (Figure~\ref{fig:interface}): A map-to-video (M2V) tool with editable masks and camera control. Participants placed masks on the map, animated them by drawing trajectories, and specified camera motion with timeline keyframes.
\end{enumerate}

\red{Both conditions were implemented in Unity and used the same ComfyUI backend, Flux1 Kontext~\cite{labs2025flux1kontextflowmatching} for reference image synthesis, and Wan2.1 VACE~\cite{jiang2025vace} for video generation. This held the generation models constant while comparing conventional image-to-video prompting with spatially grounded authoring.}

\subsubsection{\textbf{Task}}
Participants completed two video authoring tasks in both conditions:

\begin{enumerate}[leftmargin=*]
    \item \textbf{Replication Task}: We filmed three target street shots \red{(Figure~\ref{fig:replication-task})}: (1) a car arriving with a static camera, (2) a character crossing the street with a right-to-left pan, and (3) a character walking away as the camera zoomed out and tilted upward. The shots were inspired by scenes from \href{https://youtu.be/JuimqB3ofEI?si=fr5abtbk_TfpgizP}{\textit{Breakfast at Tiffany’s}}, \href{https://youtu.be/PsV1x7WAwGk?si=9HKEi0AJgzi4qRSS&t=113}{\textit{Vanilla Sky}}, and \href{https://youtu.be/rYv2a_VF328?si=5RLbiNKrPUExJvw3}{\textit{Baby Driver}}, but recreated due to limited Mapillary coverage at the original locations.
    \item \textbf{Open-Ended Task}: Participants created three clips in one location selected from ten candidates with high-quality Mapillary imagery. They could choose any theme, but were asked to maintain spatial continuity across shots. To limit study time, we allowed only one mask.
\end{enumerate}

\subsubsection{\textbf{Procedure}}
After an introduction and demographics, participants received a brief walkthrough of each system and completed both tasks in both conditions, with system order counterbalanced. After each condition, they completed questionnaires and a short semi-structured interview. Sessions lasted 90--120 minutes, and participants were compensated 25 USD/hour (75--100 USD total).

\subsubsection{\textbf{Measurement}}
For each condition, we recorded task completion time and number of iterations, and collected workload (NASA-TLX)~\cite{hart1986nasa}, usability (SUS)~\cite{brooke1996sus}, creativity support (CSI)~\cite{cherry2014quantifying}, and custom questionnaires. Participants self-rated task performance in the replication and open-ended tasks, and evaluated outputs on 7-point Likert scales for (1) spatial consistency, (2) intent alignment, (3) creative controllability, (4) ease of iteration, (5) visual quality, and (6) overall experience (Appendix Table~\ref{tab:generalQ}). They also rated \system{}-specific features on 7-point Likert scales: (1) map-based blocking, (2) trajectory design, (3) map-based camera setup, (4) framing preview, and (5) workflow adoption (Appendix Table~\ref{tab:featureQ}).

\red{Because the replication task used filmed target shots, we also measured how closely each output matched its target in both appearance and camera motion. For appearance, we compared time-aligned generated and target frames using two complementary metrics. SSIM~\cite{1284395} measures similarity in luminance, contrast, and local structure, with higher values indicating a closer match. LPIPS~\cite{8578166} measures perceptual difference using deep image features, with lower values indicating a closer match. Using both metrics allowed us to assess structural and perceptual similarity, and we averaged each metric across the frames of a clip. For camera motion, we used MegaSaM~\cite{11092772} to recover the 3D camera trajectories of the generated and target videos. We measured rotation error (RotErr) as the mean geodesic angle between their camera rotations after expressing each trajectory relative to its first frame, with lower values indicating more similar camera motion. We computed all three metrics for 72 replication clips (12 participants $\times$ 2 conditions $\times$ 3 shots), then averaged the three shots within each participant and condition to obtain 12 paired observations per metric.}

\subsection{Results}
Overall, \system{} gave participants greater control over subject placement, motion, and camera behavior in SVI-grounded video generation than the baseline. \red{Throughout, \textsuperscript{*}$p<.05$, \textsuperscript{**}$p<.01$, and \textsuperscript{***}$p<.001$.} Appendix Figure~\ref{fig:user-videos} shows that participants also produced diverse videos with \system{} in the open-ended task.

\begin{table}[h]
    \centering
    \begin{tabular}{lll}
    \toprule
    \red{\textbf{Metric}} & \red{\textbf{Baseline}} & \red{\textbf{\system{}}} \\
    \midrule
    \multicolumn{3}{l}{\red{\textit{Appearance}}} \\
    \red{\quad SSIM $\uparrow$}
        & \red{0.300 [0.289--0.306]}
        & \red{0.380 [0.378--0.391]\textsuperscript{**}} \\
    \red{\quad LPIPS $\downarrow$}
        & \red{0.758 [0.747--0.769]}
        & \red{0.732 [0.728--0.739]\textsuperscript{**}} \\
    \addlinespace
    \multicolumn{3}{l}{\red{\textit{Camera motion}}} \\
    \red{\quad RotErr (\degree) $\downarrow$}
        & \red{15.57 [15.00--15.83]}
        & \red{4.03 [3.24--5.41]\textsuperscript{**}} \\
    \bottomrule
    \end{tabular}
    \caption{\red{Similarity to the filmed targets in the replication task. Shot-averaged scores per participant, reported as median [IQR] across 12 participants (paired Wilcoxon signed-rank, Holm-corrected, $p<.01$).}}
    \label{tab:objective-similarity}
    \Description{Objective similarity between generated replication-task outputs and their filmed targets, comparing the baseline and Map2Video on three measures. SSIM and LPIPS measure per-frame appearance, and RotErr measures camera rotation error in degrees recovered with MegaSaM. Map2Video was closer to the target than the baseline on all three measures, with the largest difference in camera rotation error.}
\end{table}

\subsubsection{\textbf{Controllability and Authoring Outcomes}}

\subsubsection*{\textbf{\red{\system{} outputs were closer to the target shots, with the largest gain in camera motion.}}}
\red{Outputs authored with \system{} were closer to their targets than baseline outputs on all three metrics, and this held for every participant ($W=0$, $Z=-3.06$, $p=.0005$, Holm-adjusted $p=.0015^{**}$, $r=.88$ for each metric; Table~\ref{tab:objective-similarity}). The camera motion difference was the largest, and it appeared in all three shots: median RotErr was $0.30\degree$ vs. $4.25\degree$ for the static shot, $6.41\degree$ vs. $30.65\degree$ for the pan, and $6.18\degree$ vs. $11.16\degree$ for the zoom-out with tilt. Appearance differences were smaller and more shot-dependent: SSIM favored \system{} in all three shots, whereas per-shot LPIPS differences reached significance only in the third shot. Absolute appearance scores were modest in both conditions, which may partly reflect the resolution of the underlying street view imagery. Because both conditions used the same generation models, these differences indicate that specifying placement, motion, and camera behavior on the map and street view imagery, rather than through prompts, brought outputs closer to the intended shots.}

\begin{figure}[h]
    \centering
    \includegraphics[width=\columnwidth]{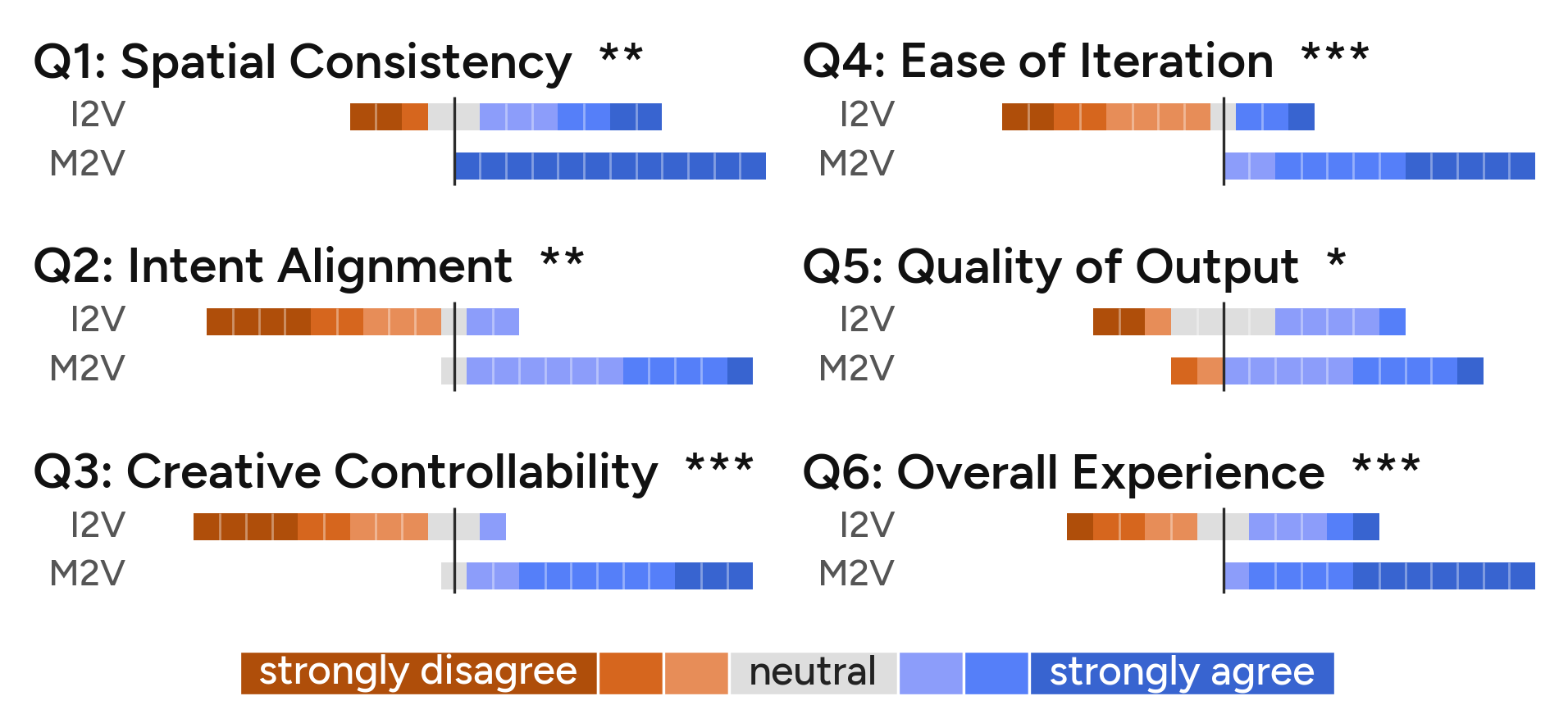}
    \caption{Custom Questionnaire Result\red{. Each square is one participant, centered on the neutral rating.}}
    \label{fig:customQ-comparison}
    \Description{Custom questionnaire results comparing Map2Video with the baseline across six items. Each bar is built from twelve cells, one per participant, coloured by the rating given and centered on the neutral rating so that disagreement extends left and agreement right. Map2Video sits further toward agreement than the baseline on all six items.}
\end{figure}

\subsubsection*{\textbf{\system{} made grounded subject and camera control more reliable.}}
\red{Figure~\ref{fig:customQ-comparison} shows the custom questionnaire results.} Creative controllability was significantly higher for \system{} ($M=5.92$, $SD=0.90$) than for the baseline ($M=2.50$, $SD=1.38$), $t(11)=6.46$, $p<.001^{***}$. This pattern was also reflected in participants’ self-evaluations. In the replication task, ratings were significantly higher for \system{} ($M=6.00$, $SD=1.13$) than for the baseline ($M=2.91$, $SD=1.78$; Welch’s $t(18.6)=-5.08$, $p<.001^{***}$). In the open-ended task, ratings were again higher for \system{} ($M=4.64$, $SD=1.23$ vs. $M=3.46$, $SD=2.06$), although the difference was not significant ($t(17.9)=-1.71$, $p=.105$).

\subsubsection*{\textbf{The baseline was easy to use, but hard to direct.}}
\begin{figure}[h]
    \centering
    \includegraphics[width=\columnwidth]{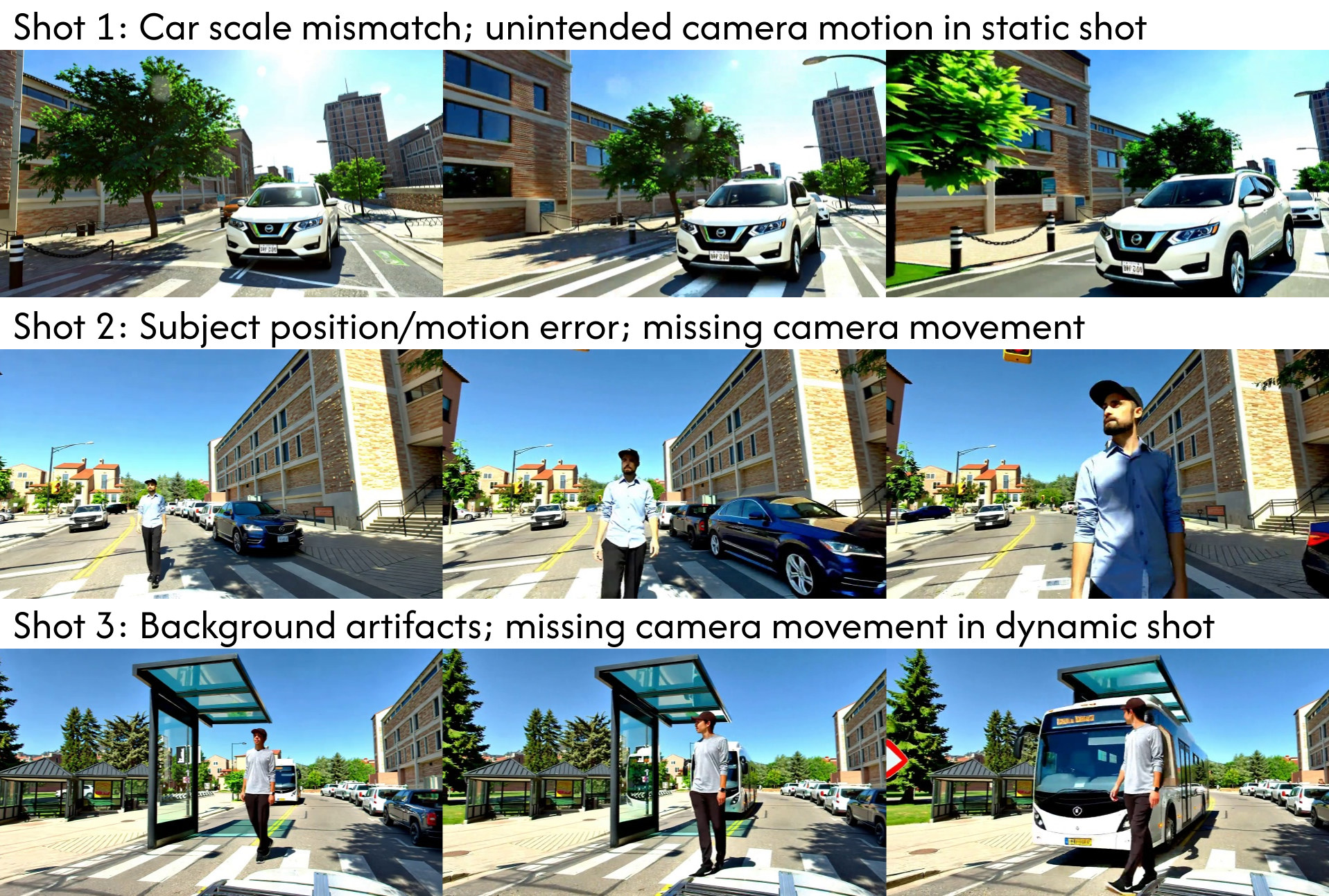}
    \caption{Baseline failure cases in the replication task.}
    \label{fig:baseline-failure-cases}
    \Description{Baseline failure cases in replication tasks, showing car scale mismatch, subject motion errors, background artifacts, and missing camera movement across shots.}
\end{figure}

In the replication task, many participants failed to reproduce intended compositions within the time limit and proceeded with imperfect references. They often had to regenerate images multiple times, yet still struggled to match subject scale, position, and motion (Figure~\ref{fig:baseline-failure-cases}). Even with detailed prompts specifying framing and camerawork, such as zoom or tilt, outputs frequently defaulted to centered subjects with minimal motion. This was consistent with our formative study (Section~\ref{section:formative}), in which filmmakers similarly described difficulty controlling composition and motion through prompt-based workflows.

\subsubsection*{\textbf{Fine-grained motion and mask control remained difficult.}}

\begin{figure}[h]
    \centering
    \includegraphics[width=\columnwidth]{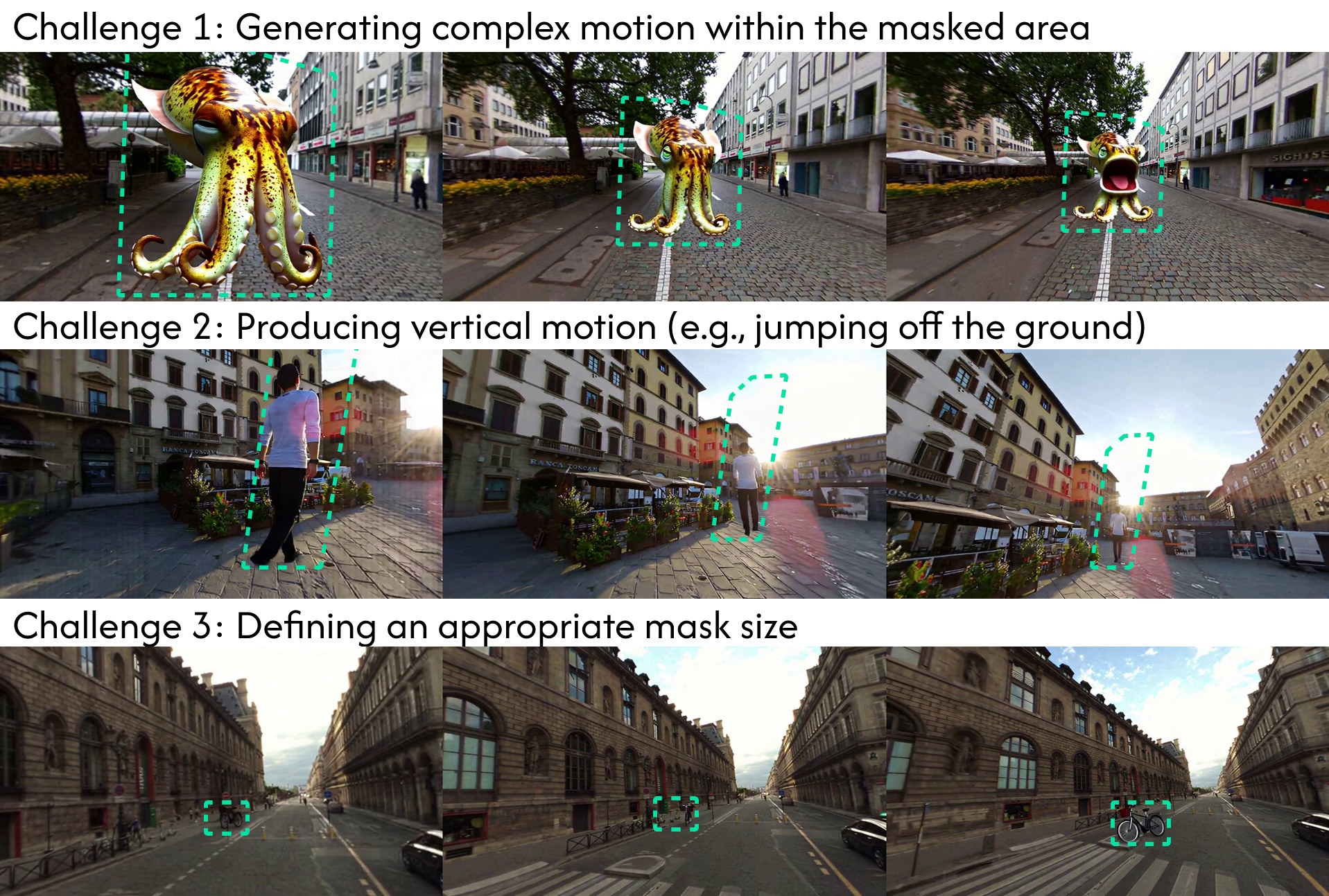}
    \caption{\system{} failure cases in the open-ended task. Green-highlighted parts are the user-defined mask. }
    \label{fig:map2video-failure-cases}
    \Description{Map2Video failure cases in open-ended tasks, showing difficulties with complex motion, vertical motion like jumping, and defining mask size.}
\end{figure}

In the open-ended task, \system{} improved overall composition control, but limitations remained in fine-grained motion and mask specification. Compound actions within a single mask, such as ``a squid rolls and explodes'' or ``a man does a somersault,'' and vertical motions such as jumping were difficult to generate reliably. Two participants (P10, P12) produced incomplete or oversized subjects, such as a bicycle without a person or a giant human (Figure~\ref{fig:map2video-failure-cases}). \red{These cases resulted from projection-based scaling's side effects and the current prototype's lack of mask-scale keyframing.} Control limitations also persisted in the baseline during Task 2, where grounded composition and motion remained difficult to direct precisely. These results suggest that \system{} shifted the main challenge from coarse scene specification to finer-grained motion and mask control.

\subsubsection{\textbf{Efficiency, Workload, and Usability}}

\subsubsection*{\textbf{\system{} reduced trial and error.}}
In Task 1, \system{} was significantly faster than the baseline (20.42$\pm$7.48 min vs. 24.65$\pm$8.14 min; Table~\ref{tab:duration_iteration}). In the baseline, participants repeatedly generated images to obtain an acceptable reference, whereas \system{} shifted effort toward layout planning and camera framing. Because video iteration counts were similar in Task 1 (baseline: 3.66$\pm$1.06; \system{}: 3.67$\pm$1.50), the efficiency gain appears to come mainly from reducing image-level trial and error rather than video refinement. In Task 2, completion time was similar across conditions, but video iterations were higher for \system{} (5.75$\pm$4.13) than for the baseline (3.68$\pm$1.33)\textsuperscript{*}. This pattern suggests that, even after scene layout and camera setup were established, fine-grained motion within the masked region still required additional iteration.

\begin{table}[t]
    \centering
    \begin{tabular}{llll}
    \toprule
    \textbf{Task} & \textbf{Item} & \textbf{Baseline} & \textbf{\system{}} \\
    \midrule
    Task 1 & Duration (min)       & 24.65 (±8.14) & 20.42 (±7.48)\textsuperscript{*} \\
           & Iteration (image)    & 8.93 (±3.82)  & NA \\
           & Iteration (video)    & 3.66 (±1.06)  & 3.67 (±1.50) \\
    \addlinespace
    Task 2 & Duration (min)       & 23.33 (±8.20) & 23.60 (±6.79) \\
           & Iteration (image)    & 5.62 (±2.27)  & NA \\
           & Iteration (video)    & 3.68 (±1.33)  & 5.75 (±4.13)\textsuperscript{*} \\
    \bottomrule
    \end{tabular}
    \caption{Task duration and iterations (paired t-test, p<.05).}
    \label{tab:duration_iteration}
    \Description{Task duration and iteration counts for the replication task and open-ended task, comparing the baseline and Map2Video. In the replication task, Map2Video showed shorter completion time than the baseline, while video iteration counts were similar. In the open-ended task, completion times were similar across conditions, but Map2Video involved more video iterations. Image iterations are reported only for the baseline. Asterisks indicate statistically significant differences.}
\end{table}

\subsubsection*{\textbf{Controllability mattered more than interface simplicity.}}
\begin{figure}[h]
    \centering
    \includegraphics[width=\columnwidth]{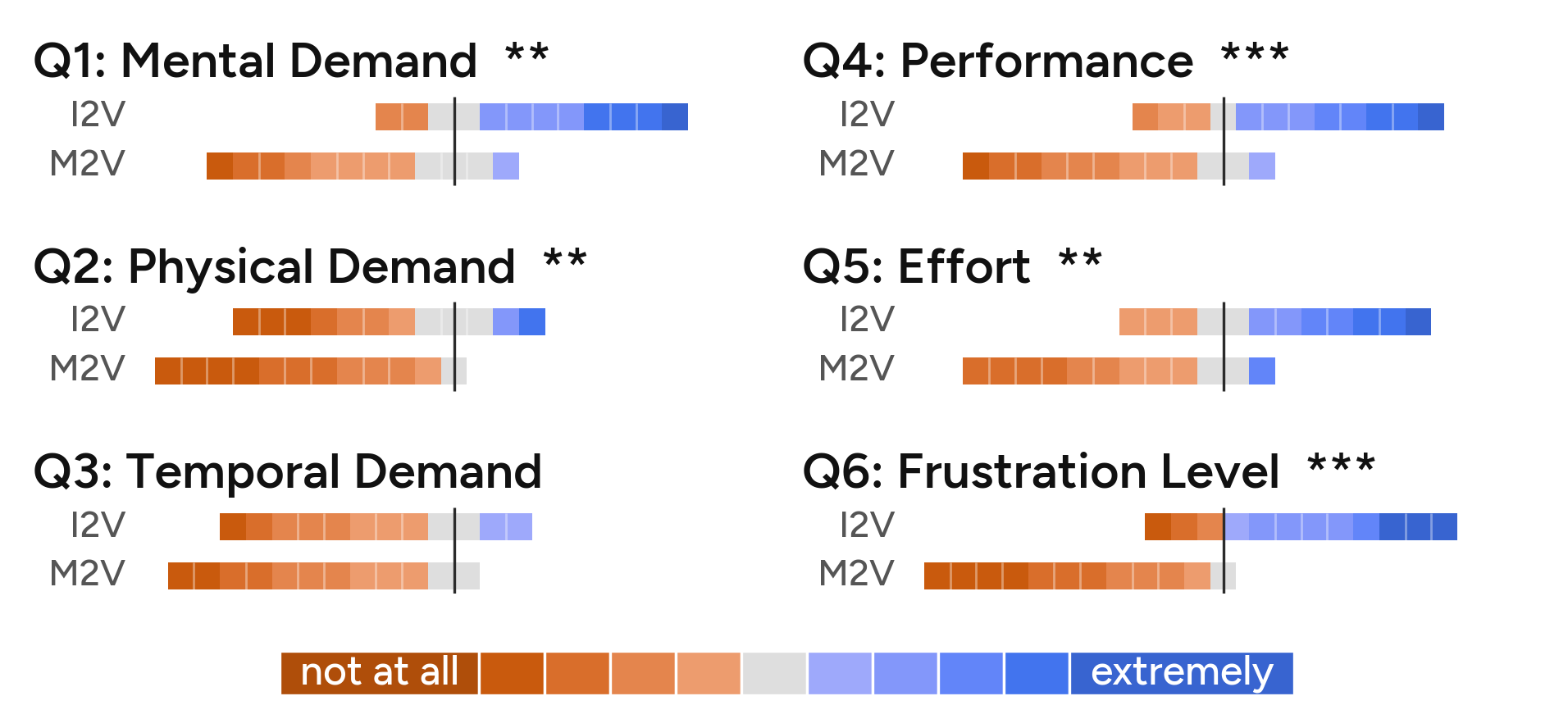}
    \caption{Task Load Questionnaire Result (NASA-TLX). The performance scale is opposite.}
    \label{fig:nasa-tlx-comparison}
    \Description{Task load results (NASA-TLX) comparing Map2Video with the baseline across six items, drawn as one cell per participant centered on the mid-point of the scale. Map2Video shows lower mental demand, physical demand, effort, and frustration, and higher self-rated performance.}
\end{figure}

\begin{figure}[h]
    \centering
    \includegraphics[width=\columnwidth]{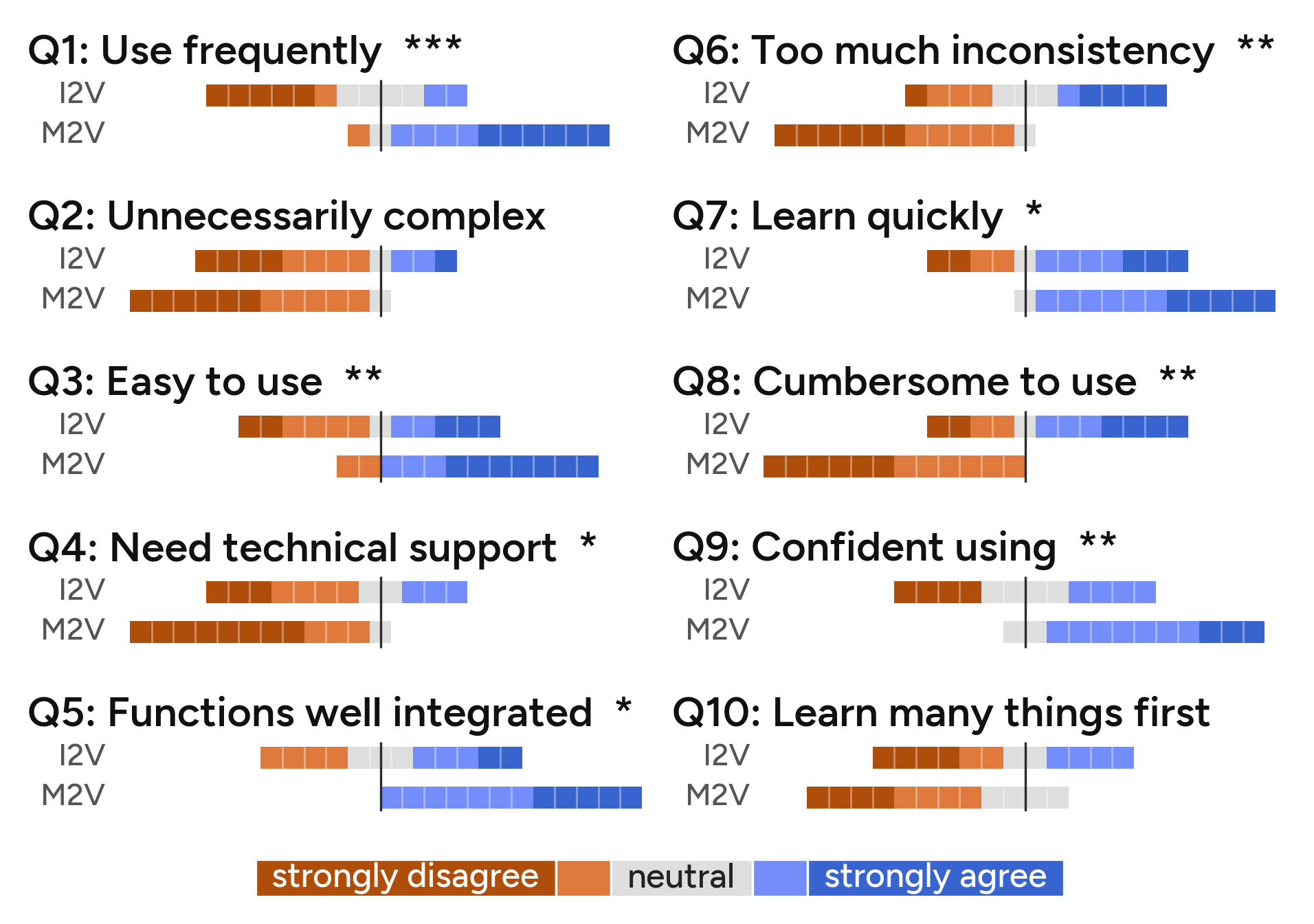}
    \caption{System Usability Scale Result (SUS).}
    \label{fig:sus-comparison}
    \Description{System Usability Scale results comparing Map2Video with the baseline across ten items, drawn as one cell per participant centered on the neutral rating. Map2Video is rated higher on ease of use, integration, quick learning, and confidence, and lower on complexity, inconsistency, and the need for technical support.}
\end{figure}

On NASA-TLX, the baseline imposed significantly higher mental demand ($t(11)=-4.01$, $p=.002^{**}$), effort ($t(11)=-4.24$, $p=.001^{**}$), and frustration ($t(11)=-4.89$, $p<.001^{***}$) (Figure~\ref{fig:nasa-tlx-comparison}). They also reported better performance with \system{} ($t(11)=-5.27$, $p<.001^{***}$). \system{} received a SUS score of 83.12 ($SD=10.43$, \textit{Excellent}), while the baseline received 51.25 ($SD=22.09$, \textit{Poor}) (Figure~\ref{fig:sus-comparison}). Despite the baseline’s simpler interface, participants rated it lower because its outputs were inconsistent and unpredictable ($W=0$, $p=.008^{**}$).

\subsubsection{\textbf{Creative Support and Output Perception}}

\subsubsection*{\textbf{\system{} made creative exploration feel more worthwhile.}}
Following prior work~\cite{cherry2014quantifying}, we computed a 5-factor weighted CSI by omitting the collaboration factor for our single-user workflow. As shown in Table~\ref{tab:CSIResult}, \system{} received a higher weighted CSI score than the baseline (76.65 vs. 44.42) and was rated higher across all factors. In particular, higher scores in Exploration and Results-Worth-Effort indicate that participants found \system{} more useful for trying different ideas and producing rewarding outcomes.

\begin{table}[h]
    \centering
    \begin{tabular}{lccc}
    \toprule
    \textbf{Factor} & \textbf{Map2Video} & \textbf{Baseline} & \textbf{Count} \\
    \midrule
    Exploration          & 16.08 (2.61) & 9.25 (6.08)  & 2.41 \\
    Enjoyment            & 16.17 (2.41) & 8.33 (5.35)  & 1.75 \\
    Expressiveness       & 15.08 (3.26) & 10.33 (6.24) & 2.00 \\
    Results Worth Effort & 14.92 (2.61) & 8.67 (6.31)  & 2.25 \\
    Immersion            & 14.17 (3.93) & 7.42 (4.78)  & 1.58 \\
    \midrule
    \textbf{Weighted Sum (CSI)} & \textbf{76.65} & \textbf{44.42} & \textbf{--} \\
    \bottomrule
    \end{tabular}
    \caption{Creative Support Index (CSI) Questionnaire Result. CSI factors (N=12) by interface with averages (SD), factor counts, and weighted sum at the bottom.}
    \label{tab:CSIResult}
    \Description{Creative Support Index questionnaire results comparing Map2Video and the baseline across five factors: exploration, enjoyment, expressiveness, results worth effort, and immersion. Map2Video received higher mean scores than the baseline on all five factors. The weighted sum was 76.65 for Map2Video and 44.42 for the baseline, indicating stronger perceived creativity support for Map2Video.}
\end{table}

\subsubsection*{\textbf{Greater control improved satisfaction and spatial consistency.}}
Overall satisfaction was higher for \system{} than for the baseline ($M=6.50$, $SD=0.67$ vs. $M=3.92$, $SD=1.78$), $t(11)=4.24$, $p<.001^{***}$ (Figure~\ref{fig:customQ-comparison}). Participants also rated \system{} as better at maintaining spatial consistency across sequential clips. As shown in Figure~\ref{fig:baseline-failure-cases}, baseline outputs often regenerated entire backgrounds when users described camera changes or referred to SVI objects such as bus stops, buildings, cars, or trees, whereas \system{} modified only masked regions. Accordingly, all participants gave \system{} the maximum spatial consistency rating, while the baseline received a lower mean rating ($M=4.42$, $SD=2.11$), $t(11)=4.24$, $p<.001^{***}$.

\subsubsection*{\textbf{Low-quality street view images and sparse viewpoints limited output quality and camera setup.}}
\begin{figure}[h]
    \centering
    \includegraphics[width=\columnwidth]{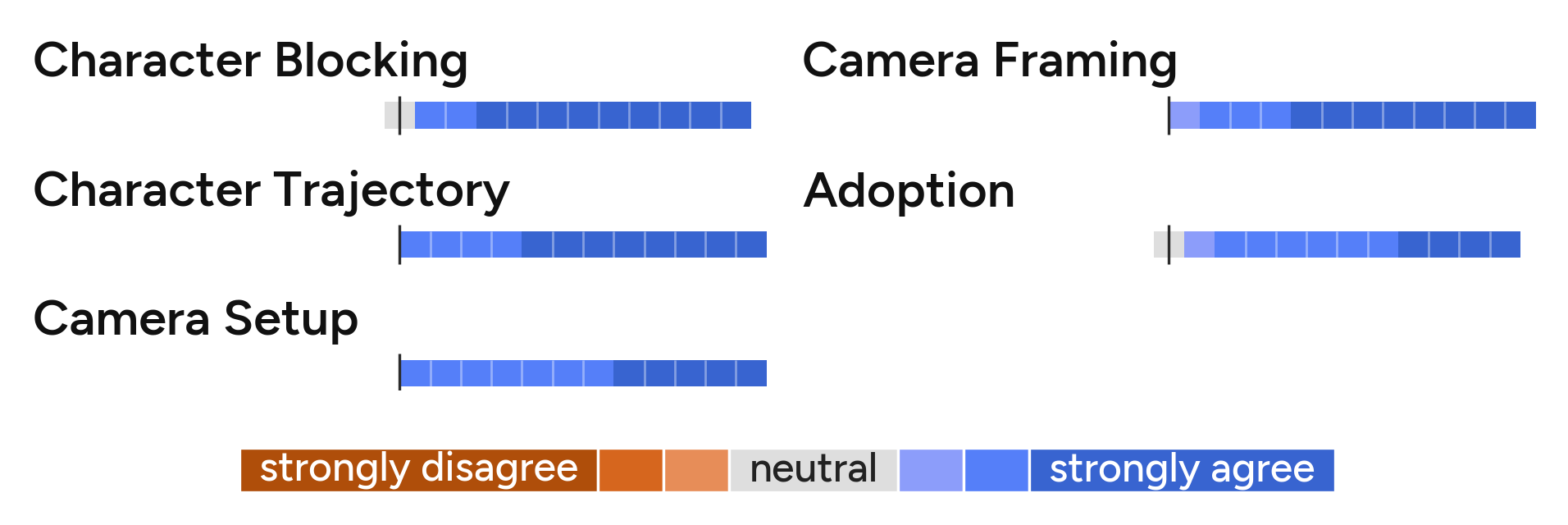}
    \caption{\system{} Feature Evaluation.}
    \label{fig:feature-evaluation}
    \Description{Feature evaluation results for Map2Video across five features, drawn as one cell per participant centered on the neutral rating. Ratings cluster in agreement for character blocking, character trajectory, camera setup, camera framing, and adoption.}
\end{figure}

Although \system{} received significantly higher output quality ratings than the baseline (Figure~\ref{fig:customQ-comparison}), its mean rating remained modest ($M=5.08$, $SD=1.38$). Even after upscaling, Mapillary backgrounds still contained degraded pixels, affecting subsequent video quality. The same source limitations appeared in feature-level ratings; trajectory drawing received the highest score ($M=6.67$, $SD=0.49$), while map-based camera setup received the lowest ($M=6.42$, $SD=0.51$) (Figure~\ref{fig:feature-evaluation}). This lower rating likely reflects Mapillary’s limited 360\degree{} viewpoints, which restricted precise street view selection.

\subsection{Qualitative Insights}

\subsubsection*{\textbf{Street View Imagery as a Grounded Starting Point}}
Participants described street view imagery as a grounded starting point for ideation, rather than merely a reference source. Unlike text prompting, SVI provided spatial context that helped them make creative decisions about plot, action, and shot design. For example, P2 noted that maps helped ``channel creativity into the plot and action,'' while P11 said that beginning with an existing environment was ``better than a blank canvas.'' Others used it to think through continuity and pre-production details such as building shadows (P8). At the same time, several participants noted that fixed real-world backgrounds could constrain more surreal or imaginative directions, suggesting a tradeoff between grounding and abstraction.

\subsubsection*{\textbf{Balancing Control and Randomness}}
Participants valued both deliberate control and occasional randomness, suggesting that they supported different creative modes. \system{} made trajectories and camera motion more intentional and predictable, reducing guesswork during authoring. At the same time, several participants saw value in the baseline’s unpredictability. P9 noted that it ``made me try more variations,'' suggesting that randomness could support improvisational exploration. These reflections suggest an opportunity to combine grounded control with room for surprise and divergence.

\subsubsection*{\textbf{Agency Through Preview and Feedback}}
Participants often described \system{} as closer to ``directing'' than ``generating.'' They valued being able to place masks directly within the camera view, preview trajectories and camera paths, and decide whether to proceed before full generation. This preview-feedback loop contributed to a stronger sense of agency. P1 noted that manually setting trajectories meant ``less time spent waiting,'' while P10 emphasized that previewing masks reduced unnecessary iterations. Similarly, P4 described ``giving up faster'' with the baseline because too much effort was spent on uncertain generations.

\subsubsection*{\textbf{Expanding Camera and Motion Control}}
Participants pointed to several directions for extending grounded control beyond the current prototype. Some requested more flexible camera control beyond fixed $360\degree$ image points, for example by sketching camera trajectories directly on the map or SVI to achieve bird’s-eye views or dolly-like shots (P4, P11). Others wanted more expressive subject motion, including vertical trajectories and shape changes for actions such as flying, exploding, or transforming. They also proposed tighter integration with familiar workflows, such as Premiere-style shortcuts (P1, P3) and Blender export for timeline editing (P5), along with depth recognition (P5, P12) and object interaction (P10) through 3D reconstruction of SVI environments and characters.\\

Overall, these insights suggest that \system{} supported grounded video generation through greater control, more concrete planning, and iterative feedback. They also highlight opportunities for future tools to support more expressive motion, flexible camera control, and occasional departures from real-world constraints.

\section{Limitations and Future Work}
\label{sec:discussion}

\subsubsection*{\textbf{Addressing Street View Source Limitations}}
Some limitations stemmed not from the interface, but from the underlying street view source. Fixed eye-level viewpoints constrained camera control, while low-quality Mapillary imagery reduced output quality even after upscaling. \red{SVI also constrains the initial background composition through the available viewpoint, lighting, bystanders, and obstacles. Although our workflow can remove unwanted pedestrians and upscale low-resolution imagery, it does not yet provide full control over background composition or appearance.} Although street view imagery served as a useful grounded scaffold, future systems could better separate spatial grounding from source appearance, support smoother viewpoint changes beyond sparse, discrete views, and allow changes in appearance such as atmosphere or time of day while preserving scene geometry. Possible directions include panoramic environment reconstruction~\cite{wang2024perf}, autoregressive video diffusion over street-view sequences~\cite{deng2024streetscapes}, \red{novel-view synthesis along flexible camera trajectories~\cite{yan2025streetcrafter},} and layout-conditioned generation approaches that preserve scene geometry while allowing controllable changes in appearance. More broadly, while our system focuses on street view imagery and maps, the underlying idea of controllable grounded authoring could extend to other structured scene priors, such as 3D models~\cite{polysphere2025paris} and point clouds~\cite{zheng2026versecrafter}.

\subsubsection*{\textbf{Supporting Fine-Grained Motion Control}}
\red{Fine-grained motion remained difficult to specify, especially for compound actions, vertical movement, and changes in mask scale. Although mask position and projected size respond to the map and SVI geometry, the current prototype does not support keyframing mask scale, shape, or elevation over time.} Planning motion on the map also did not always translate intuitively to street view-based editing. Future work could focus on improving motion specification and preview across both map and SVI views. One direction is to support multi-scale authoring, where users move between high-level geographic planning and fine-grained street-level adjustment. \red{Keyframes for mask scale, shape, and elevation could provide more direct control over actions such as approaching the camera, jumping, or morphing.} Another is to provide motion preview through real-time text-to-motion models~\cite{zhao2024dartcontrol, zhao2026ardy}. For example, text annotations attached to character trajectories could allow users to specify actions or movement styles that update interactively during preview.

\red{\subsubsection*{\textbf{Extending Grounded Authoring Beyond Filmmaking}}
Although \system{} is designed for filmmaking previsualization, its core workflow could support other tasks that involve communicating how events or proposed changes may unfold in specific real-world locations. For example, accident reconstruction teams could visualize hypothesized sequences of vehicle and pedestrian movement to support discussion among investigators or stakeholders. Safety trainers could present location-specific scenarios, such as a cyclist entering a blind intersection, when explaining potential risks or proposed procedures before an on-site exercise. Urban planners could similarly create situated visualizations of proposed changes to crossings, traffic flow, or public-space use for stakeholder consultation. Across these settings, maps and SVI provide shared spatial context, while editable trajectories and camera framing help turn abstract scenarios into concrete, location-grounded visualizations.}

\subsubsection*{\textbf{Ethical Considerations}}
In developing \system{}, we recognize ethical implications in using generative AI together with street view imagery for filmmaking. Because the system relies on real-world visual data, it raises concerns around privacy, data provenance, and geographic coverage bias, which may affect what locations can be represented faithfully or controllably. \red{Generated depictions of real locations may also be mistaken for events that actually occurred there.} Our goal is not to automate storytelling, but to support creators in authoring spatially grounded scenes with greater control. To mitigate these risks, users remain responsible for selecting locations, specifying prompts, and defining motion and camera behavior, keeping AI in an assistive rather than autonomous role. \red{Future systems should preserve source attribution and licensing information, clearly distinguish generated content from captured imagery, and support the removal or anonymization of identifiable bystanders.}

\section{Conclusion}
We introduced \system{}, a street view imagery-driven system for more controllable SVI-grounded video authoring. By combining map-based annotation with video inpainting, \system{} helped users author location-specific scenes with more direct control over subject placement, motion, and camera behavior. Our studies showed that street view imagery served as a grounded starting point, while the rapid preview-feedback loop supported iterative refinement. These findings suggest opportunities for future systems to expand camera and motion control, improve coordination between map and street view representations, and reduce limitations of current street view imagery sources.

\bibliographystyle{ACM-Reference-Format}
\bibliography{references}

\clearpage

\appendix
\section*{Appendix}\label{section:appendix}
\noindent

\subsection*{Survey on AI Filmmaking}\label{subsection:survey}
As generative AI video tools enter filmmaking practice, their real-world use and control challenges remain underexplored. To inform our system design, we conducted a formative survey with AI-assisted filmmaking practitioners.

\subsubsection*{\textbf{Method}}

The survey examined usage patterns, workflows, and challenges. It was distributed through online AI filmmaking communities (Creative Refuge, Facebook groups, and Discord). The survey included questions about the AI video tools participants used, their use cases, and the challenges they encountered. Participation was voluntary and uncompensated, though respondents were informed about the possibility of a compensated follow-up formative study.

In total, 34 filmmakers (7 female, 27 male) completed the survey. All had experience with AI-assisted filmmaking, and 30 of 32 respondents also had traditional filmmaking experience. Participants held roles such as director, screenwriter, producer, editor, and cinematographer. Many worked across the full AI production pipeline, including prompt design, creative direction, and editing.

Most participants (29/34) used multiple AI tools, with self-assessed proficiency skewing advanced (18 advanced, 5 intermediate, 2 beginner, 9 unspecified). Table~\ref{tab:ai_video_tools} lists the reported tools.

\begin{table}[h]
  \centering
  \begin{tabular}{@{}ll|ll@{}}
    \toprule
    Tool Name & Mention & Tool Name & Mention \\
    \midrule
    Runway              & 29 & Luma / Lumalabs   & 2  \\
    Kling               & 18 & Seedance          & 2  \\
    Veo / Veo 3         & 12 & Minimax           & 2  \\
    ComfyUI             & 11 & Reve              & 1  \\
    Pika                & 6  & Pixverse          & 1  \\
    Sora                & 5  & Kaiber            & 1  \\
    Flux / Flux Kontext & 3  & Stable Video Diffusion & 1 \\
    Vidu                & 1  & Moonvally         & 1  \\
    Dreamina            & 1  & RunningHub        & 1  \\
    \bottomrule
  \end{tabular}
  \caption{AI video generation tools reported by participants.}
  \label{tab:ai_video_tools}
  \Description{AI video generation tools reported by participants in the formative survey, with the number of mentions for each tool. Runway was mentioned most often, followed by Kling, Veo or Veo 3, and ComfyUI. The remaining tools were mentioned less frequently.}
\end{table}

\subsubsection*{\textbf{Current Practices}}
Filmmakers reported using generative AI video tools across all stages of the filmmaking process: pre-production, production, and post-production, with varying degrees of integration into traditional workflows. In pre-production, many participants used tools like \textit{Midjourney} and \textit{Runway} to create mood boards, storyboards, and previsualizations, allowing for rapid iteration on visual concepts before committing to full production. During production, several participants used tools such as \textit{Veo}, \textit{Kling}, and \textit{ComfyUI} to generate video sequences, especially for stylized or speculative scenes that would be difficult or costly to produce conventionally. Some used AI-generated clips as B-roll or animation to complement live-action footage. In post-production, AI tools were commonly applied to editing, compositing, background replacement, stylization, and visual effects. While a few participants (6) used AI tools to support traditional workflows, the majority reported \textit{``producing fully AI-generated films''}, handling scripting, scene design, and video generation with AI tools, and refining the output using conventional editing software such as Premiere Pro or DaVinci Resolve. In these cases, AI filmmaking either replaced conventional production processes entirely or filled specific gaps such as location footage, animation, or VFX. Participants reported \textit{``creative exploration''}, \textit{``faster turnaround''}, and the ability to \textit{``visualize ideas that would be too expensive or impossible to shoot''} as key motivations for adopting generative AI tools. Participants implied the use of diverse strategies to prepare reference images for video generation, including sourcing real-world images from stock platforms (e.g., Getty), generating visuals through image-creation tools, and constructing virtual environments with 3D/virtual set software such as Unreal Engine. %
Among the portfolio examples shared by participants, the majority focused on relatively static single-subject shots, often with shallow depth of field or blurred backgrounds. We also observed examples centered on vehicle movement through urban environments.

\subsubsection*{\textbf{Challenges}}
One of the most common challenges reported by participants was \textit{\textbf{``maintaining visual and spatial consistency''}} (mentioned by 17 respondents). These issues included inconsistencies in character appearance, lighting, backgrounds, and overall scene geography, especially across longer sequences. Nine participants noted difficulties with \textit{\textbf{``controlling composition and camera movement''}}, such as aligning shots with intended framing or achieving smooth transitions while the character is moving. \textit{\textbf{``Prompt adherence''}} was another frequent concern, with 12 participants describing frustration when generated outputs did not match their intended visual or narrative details. This included problems with \textit{\textbf{``character placement''}} (11) and specifying \textit{\textbf{``character actions''}} via text (3). \textit{\textbf{``Quality degradation over time,''}} such as hallucinated elements and inconsistent lip-sync and gesture timing, was mentioned by 6 participants. \textit{\textbf{``Multi-character coordination''}} was also noted as a difficulty by 6 participants. Together, these responses point to broader difficulties in controlling generated videos to match filmmakers' intentions. In addition to generation issues, participants pointed to broader obstacles: 8 mentioned the \textit{\textbf{``high cost''}} of using certain AI tools, 5 reported the \textit{\textbf{``steep learning curve''}} in tools like \textit{ComfyUI}, and 4 noted the \textit{\textbf{``difficulty of keeping up with rapidly evolving technologies''}}. Although some participants observed improvements in character consistency and tool capabilities, most agreed that achieving coherent, high-quality outputs still requires extensive iteration, manual correction, and post-production work. These responses, together with the submitted portfolio examples, suggested a promising opportunity for more directly controlling subject placement, motion, and camera behavior in location-grounded scenes.

\begin{figure*}[h]
    \centering
    \includegraphics[width=\textwidth]{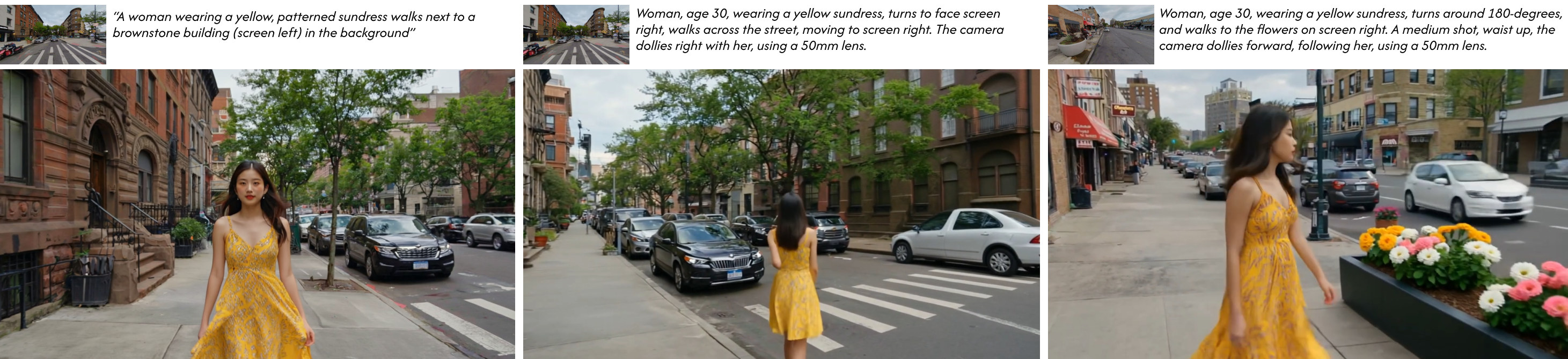}
    \caption{An example of three sequential user-created shots and corresponding input street view images and user prompts.}
    \label{fig:formative-user-artifact-example}
    \Description{Formative study output example showing a three-shot sequence: a woman walks down the street, turns right to cross it, and then turns toward flowers along the sidewalk.}
\end{figure*}

\begin{table*}[t]
    \centering
    \renewcommand{\arraystretch}{1.2}
    \setlength{\tabcolsep}{6pt}
    \begin{tabular}{p{0.18\textwidth} | p{0.34\textwidth} | >{\centering\arraybackslash}m{0.38\textwidth}}
    \toprule
    \textbf{Step} & \textbf{Description} & \textbf{Formula} \\
    \midrule

    1. Local ENU projection &
    Convert subject geodetic coordinates $(\varphi_a,\lambda_a)$ into local East--North--Up (ENU) offsets relative to the camera $(\varphi_c,\lambda_c)$. &
    $\Delta x \approx R\cos\varphi_c(\lambda_a-\lambda_c), \quad
    \Delta z \approx R(\varphi_a-\varphi_c)$
    \\
    \midrule

    2. Range and bearing &
    Compute ground distance $d$ and bearing angle $\psi_{ca}$ between subject and camera. &
    $d = \sqrt{\Delta x^2+\Delta z^2}, \quad
    \psi_{ca} = \operatorname{atan2}(\Delta x,\Delta z)$
    \\
    \midrule

    3. Camera-relative angles &
    Subtract camera heading $\psi_c$ and pitch $\theta_c$ to obtain relative azimuth and elevation. &
    $\Delta\psi = \psi_{ca}-\psi_c, \quad
    \Delta\theta = -\operatorname{atan2}(h_c,d) - \theta_c$
    \\
    \midrule

    4. Pinhole projection &
    Map relative angles into normalized screen coordinates using a pinhole camera model. &
    $s_x = \frac{\tan(\Delta\psi)}{\tan(\alpha_h/2)}, \quad
    s_y = -\frac{\tan(\Delta\theta)}{\tan(\alpha_v/2)}$
    \\
    \midrule

    5. Screen scaling &
    Convert normalized coordinates to pixel coordinates $(u,v)$ given screen resolution $(W,H)$. &
    $u = \frac{s_x+1}{2}W, \quad
    v = \frac{s_y+1}{2}H$
    \\

    \bottomrule
    \end{tabular}
    \caption{Subject placement from geodetic coordinates to panoramic screen coordinates.}
    \label{tab:actorplacement}
    \Description{Five-step procedure for converting subject geodetic coordinates into panoramic screen coordinates. The table lists each step, a brief description, and the corresponding formula, including local ENU projection, range and bearing computation, camera-relative angle calculation, pinhole projection, and screen scaling. Together, these steps define how Map2Video places subjects on a panoramic street-view image.}
\end{table*}

\begin{table*}[t]
    \centering
    \begin{tabularx}{\textwidth}{l | X}
    \toprule
    \textbf{Dimension} & \textbf{Question Item} \\
    \midrule
    Task 1 Self-Evaluation & I think I did well in the replication task, and the quality of the output meets my expectations. \\
    Task 2 Self-Evaluation & I think I did well in the open-ended task, and the quality of the output meets my expectations. \\
    Spatial Consistency & The three clips appeared consistent as if filmed in the same location. \\
    Intent Alignment & The output reflected my intended direction well. \\
    Creative Controllability & It was easy to control the system to produce the changes I wanted. \\
    Ease of Iteration & I could quickly try, modify, and refine my ideas. \\
    Quality of Output & The generated results met my expectations for visual quality. \\
    Overall Experience & Overall, my experience with the system was positive. \\
    \bottomrule
    \end{tabularx}
    \caption{Custom questionnaire items used for self-evaluation and output assessment. All items were rated on a 7-point Likert scale (1 = Strongly Disagree, 7 = Strongly Agree).}
    \label{tab:generalQ}
    \Description{Custom questionnaire items used for self-evaluation and output assessment. The table lists seven dimensions: self-evaluation for Task 1 and Task 2, spatial consistency, intent alignment, creative controllability, ease of iteration, quality of output, and overall experience. Each row provides the statement participants rated on a 7-point Likert scale.}
\end{table*}

\begin{table*}[t]
    \centering
    \begin{tabularx}{\textwidth}{l | X}
    \toprule
    \textbf{Feature} & \textbf{Question Item} \\
    \midrule
    Character Blocking & Placing characters on the map helped me maintain spatial continuity. \\
    Character Trajectory & Designing character trajectories felt intuitive and useful. \\
    Camera Setup & Setting up cameras on the map helped me visualize spatial relationships between shots. \\
    Camera Framing & Previewing framing helped me plan the shot effectively. \\
    Adoption & I would incorporate these features into my creative workflow. \\
    \bottomrule
    \end{tabularx}
    \caption{Feature evaluation questionnaire items. All items were rated on a 7-point Likert scale (1 = Strongly Disagree, 7 = Strongly Agree).}
    \label{tab:featureQ}
    \Description{Feature evaluation questionnaire items for Map2Video. The table lists five features--character blocking, character trajectory, camera setup, camera framing, and adoption--and the statement used to evaluate each one. The items assess whether these features helped participants maintain spatial continuity, design motion, visualize relationships between shots, plan framing, and consider adopting the system in their creative workflow.}
\end{table*}

\begin{figure*}[h]
    \centering
    \includegraphics[width=\textwidth]{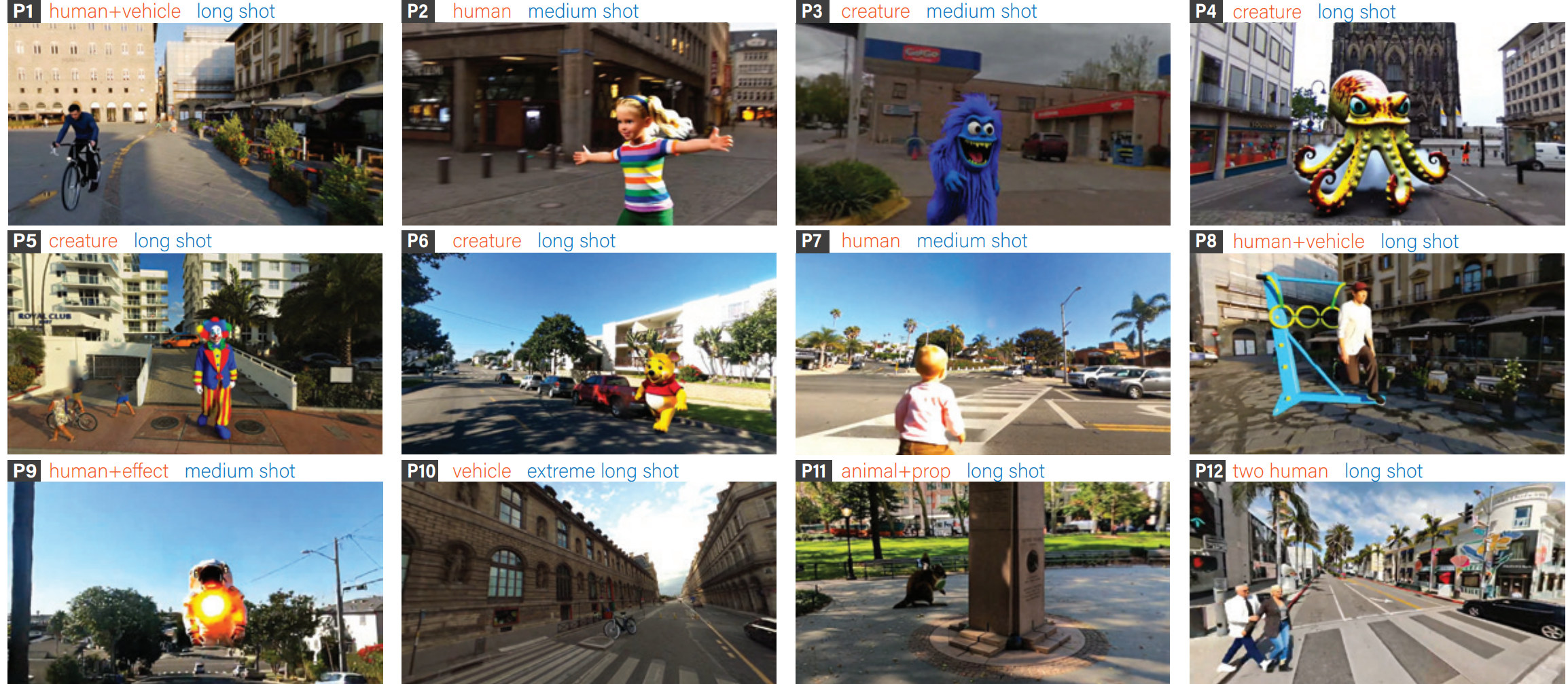}
    \caption{Examples of video artifacts created by users with the \system{} system, illustrating diverse variations in
\textcolor{Bittersweet}{subjects}, \textcolor{RoyalBlue}{shot styles}, geographic places, and subject positioning within the frame.}
    \label{fig:user-videos}
    \Description{User-created video artifacts showing variations in subjects, creatures, props, shot styles, locations, and framing across twelve panels.}
\end{figure*}

\end{document}